\documentclass[preprintnumbers,groupedaddress,nofootinbib,aps]{revtex4}
\pdfoutput=1
\usepackage{multirow,array}
\usepackage{amsmath,amssymb}
\usepackage{graphicx}
\usepackage{color}
\usepackage{url}
\usepackage{feynmp}
\usepackage[alsoload=hep]{siunitx}
\usepackage[pdfborder={0 0 0}]{hyperref}
\usepackage[utf8]{inputenc}
\usepackage{xspace}
\usepackage{setspace}
\usepackage{threeparttable}
\usepackage{relsize}
\usepackage{ifthen}
\usepackage{lineno}
\usepackage{enumerate}
\usepackage{makeidx}
\usepackage{showidx}

\newcommand{\bea}{\begin{align}}
\newcommand{\eea}{\end{align}}
\newcommand{\beq}{\begin{equation}}
\newcommand{\eeq}{\end{equation}}

\newcommand{\bs}{\begin{split}}
\newcommand{\es}{\end{split}}

\newcommand{\bi}{\begin{itemize}}
\newcommand{\ei}{\end{itemize}}
\newcommand{\bc}{\begin{center}}
\newcommand{\ec}{\end{center}}
\newcommand{\bac}{\begin{array}{c}}
\newcommand{\bacc}{\begin{array}{cc}}
\newcommand{\ea}{\end{array}}

\def\spa#1.#2{\langle#1\,#2\rangle}
\def\spb#1.#2{[#1\,#2]}

\newcommand{\UGeV}{\ensuremath{\mathrm{GeV}}\xspace}

\newcommand{\CF}{\ensuremath{C_F}}
\newcommand{\CA}{\ensuremath{C_A}}

\newcommand{\sla}[1]{\ensuremath{{#1\kern-0.45em/}}}

\newcommand{\eg}{{\it e.g.}\xspace}

\newcommand\cdf{CDF\xspace}
\newcommand\CDF{\cdf}

\newcommand\lep{L\scalebox{0.8}{EP}\xspace}

\newcommand\LEP{\lep}

\newcommand\LHC{L\protect\scalebox{0.8}{HC}\xspace}
\newcommand\ATLAS{\atlas}
\newcommand\atlas{A\protect\scalebox{0.8}{TLAS}\xspace}
\newcommand\CMS{\Cms}
\newcommand\Cms{C\protect\scalebox{0.8}{MS}\xspace}

\newcommand{\MCatNLO}{M\protect\scalebox{0.8}{C}@N\protect\scalebox{0.8}{LO}\xspace}

\newcommand{\MadGraphFive}{M\protect\scalebox{0.8}{AD}G\protect\scalebox{0.8}{RAPH}5\xspace}

\newcommand{\Herwig}{H\protect\scalebox{0.8}{ERWIG}\xspace}

\newcommand\herwigpp{\Herwig\plusplus\xspace}
\newcommand{\HerwigPP}{\herwigpp}

\newcommand{\Pythia}{P\protect\scalebox{0.8}{YTHIA}\xspace}

\newcommand\PythiaEight{\Pythia~8\xspace}

\newcommand{\OpenLoops}{O\protect\scalebox{0.8}{PEN}L\protect\scalebox{0.8}{OOPS}}

\newcommand{\Sherpa}{S\protect\scalebox{0.8}{HERPA}\xspace}

\newcommand{\Rivet}{R\protect\scalebox{0.8}{IVET}\xspace}

\newcommand{\Fastjet}{F\protect\scalebox{0.8}{AST}J\protect\scalebox{0.8}{ET}\xspace}

\DeclareRobustCommand{\plusplus}{\raisebox{0.2ex}{\smaller++}}

\newcommand\lapproxeq{\lower .7ex\hbox{$\;\stackrel{\textstyle <}{\sim}\;$}}
\newcommand\gapproxeq{\lower .7ex\hbox{$\;\stackrel{\textstyle >}{\sim}\;$}}

\hypersetup{
  pdfauthor={Dorival Goncalves, Frank Krauss, Robin Linten},
  pdftitle={Distinguishing b-quark and gluon jets with a tagged b-hadron},
  pdfkeywords={QCD, Jets, bottom quarks, heavy flavour, tagging}
}
\begin{document}
\title{Distinguishing $b$-quark and gluon jets with a tagged $b$-hadron}

\author{Dorival Gon\c{c}alves, Frank Krauss, Robin Linten}

\affiliation{Institute for Particle Physics Phenomenology\\
  Physics Department, Durham University\\
  Durham DH1 3LE, United Kingdom}

\begin{abstract}
  \noindent
  Based on the knowledge of the QCD radiation pattern, observables to distinguish 
  jets containing one and two $b$-hadrons are discussed. A simple method is used to
  combine pairs of the most sensitive observables, girth, number of charged tracks
  and the energy or momentum fraction of the leading $b$-hadron with respect to the
  jet, into one discriminator.  Their efficiencies, on particle level, are estimated
  and found to improve the performance and the robustness of the observables in
  different momentum slices.
\end{abstract}

\preprint{IPPP/15/75}
\preprint{DCPT/15/150}  
\preprint{MCNET-15-36}
\maketitle

\section{Introduction}
Jets containing a bottom quark play a significant role in many analyses at the
\LHC, both in searches for new physics and in further studies of the Standard
Model (SM).

As an illustrative example, consider the  measurements of the
phenomenologically relevant Yukawa couplings of the newly found Higgs boson
to quarks of the third generation, top and bottom quarks.  One of the 
processes central to this measurement  is the production
of a Higgs boson in association with top-quarks, $pp\to t\bar{t}H$, where the
Higgs boson decays into a $b\bar{b}$ pair.  For this study, both the signal
and the dominant background processes are understood at next-to leading order
in QCD~\cite{Beenakker:2001rj,Beenakker:2002nc,Dawson:2002tg,
Dawson:2003zu, Bredenstein:2009aj,Bredenstein:2010rs}.
More modern fixed-order calculations, performed with automated tools such as
\OpenLoops+\Sherpa~\cite{Cascioli:2011va,Gleisberg:2003xi,Gleisberg:2008ta}
or \MadGraphFive~\cite{Alwall:2011uj,Alwall:2014hca}, have successfully been
embedded in hadron-level simulations based on \MCatNLO~\cite{Frixione:2002ik},
for the signal process $t\bar{t}H$~\cite{Frederix:2011zi}, and the dominant
irreducible background $t\bar{t}b\bar{b}$~\cite{Cascioli:2013era}.
Multijet merging technologies at NLO~\cite{Hoeche:2012yf,Frederix:2012ps,
  Lonnblad:2012ng,Platzer:2012bs}
have successfully been applied to the production of top--\-anti-top pairs in
conjunction with jets~\cite{Hoeche:2013mua,Hoeche:2014qda}, thereby also
providing a handle on this background.  Combined, this work represents an
amazing technological development.  However, looking at the analysis strategy
employed by both \ATLAS and \CMS, it becomes clear that the experimental
cuts shape the background and the signal to look relatively similar, rendering
them hard to distinguish.  In the end it essentially reduces this analysis
to the counting of events with a suitable number of $b$-jets -- 3 or 4 --
within certain acceptance regions~\cite{Khachatryan:2015ila,Aad:2015gra,
Moretti:2015vaa,Buckley:2015vsa}.

One of the problems arising from this kind of analysis is related to the fact
that they rely on the identification of $b$-quarks through jets with a $b$-tag.
This identification is realised by $b$-tagging conditions~\cite{Aad:2015ydr,
ATLAS-CONF-2011-102,1748-0221-8-04-P04013}.
Examples include criteria based on displaced vertices with a certain impact
parameter significance, the presence of soft muons inside the jet, which may
stem from such a displaced vertex, or criteria based on the further decay
chain and their possible impact on the intrinsic shape of the such tagged
jet~\cite{1742-6596-119-3-032032}.
Usually, the acceptance rate of jets including a $b$-hadron based on such tags
is relatively high, between 60\% and 70\%, while the rejection rate of light
jets containing no such hadron reaches well beyond 90\% at typical working
points.  However, this simple tagging technology may fail to reliably
identify jets containing two $b$-hadrons, which can originate from a
$g\to b\bar{b}$ splitting.  This translates into limitations in distinguishing 
``legitimate'' $b$-jets stemming from a $b$-quark from gluon (or other light)
jets, thereby hampering analyses of processes with $b$'s produced in the hard
interaction.  This is further exarcerbated by the absence of very precise
theory estimates of the gluon splitting: its description by the parton shower
is possibly not quite as reliable as one would naively assume.  Earlier
analyses by the \LEP collaborations measured this splitting probability with
large statistical and systematic errors in the range of
$(0.21\%$-$0.31\%)\pm 0.1\%$, while the parton shower programs
usually arrived at rates of just below $0.2\%$~\cite{Abreu:1997nf,
  Barate:1998vs,Abreu:1999qh,Abbiendi:2000zt}.  This immediately translates
into the need not only to measure the $g\to b\bar{b}$ transition such that
the modern parton shower algorithms can be compared and, if necessary,
improved through direct comparison.  It also motivates to construct robust
and reliable observables discriminating the ``real'' $b$-jets from those jets
where $b\bar{b}$ pair emerges from gluon splitting.

Some early attempts at this identification were performed by 
\CDF~\cite{Acosta:2004nj} by trying to identify two secondary vertices 
in the jet consistent with two $b$-hadrons from a sample of already tagged
events. Both \ATLAS~\cite{ATLAS-CONF-2012-100} and \CMS~\cite{CMS-DP-2015-038}
are also working on this identification, with varying levels of success. 
Due to the intrinsic difficulty of finding two separate secondary vertices 
belonging to $b$-hadrons, these searches are typically using observables
related to the jet and the vertex. Both collaborations use sophisticated
multivariate analysis tools to define their discriminators. 

This short letter aims to further explore the very same problem.  Using
well-established features in the QCD radiation pattern and simple geometric
considerations motivates to use a combination of jet shapes and secondary
vertex finding to distinguish $b$-jets from what will be called $b\bar{b}$-jets 
in the rest of the paper. This paper is organised as follows:  In Section~\ref{sec:observables}
the most sensitive jet shape observables are reviewed and possible improvements when 
combining them with a reconstruction of fragmentation function observables are discussed. 
The analysis is performed in Section~\ref{sec:analysis}, the results presented in Section~\ref{sec:combination}
and the summary in Section~\ref{sec:summary}.

\section{Shaping b-jets: Kinematic observables}
\label{sec:observables}

It is well-known that the fragmentation function $F(x)$ of $b$-quarks, is
relatively hard, peaking close to $x\approx 1$.   Here, $x$ denotes the
$b$-hadron energy or momentum fractions $x_E$ or $x_p$ with respect to the
underlying $b$-quark jet.  This behaviour is due to the fact that the finite
masses of the $b$-quarks shield the collinear divergence in gluon emissions
off the quark, thereby effectively suppressing the emission of energetic
secondary partons, a phenomenon sometimes called ``dead cone effect''.  As
a result, $b$-quarks tend to retain most of their energy -- in contrast to
light partons -- and thus the $b$-hadrons more or less have energies and
momenta very similar to the $b$-quark when it was produced in the hard
process.  Conversely, $b$-quarks originating from a gluon splitting tend to
have a fairly symmetric share in the energy of the original gluon, which
they retain during fragmentation.  As a result the emerging $b$-hadrons, and
in particular also the harder of the two, tend to have an energy fractions
well below unity.
\begin{figure}[b!]
  \includegraphics[scale=0.48]{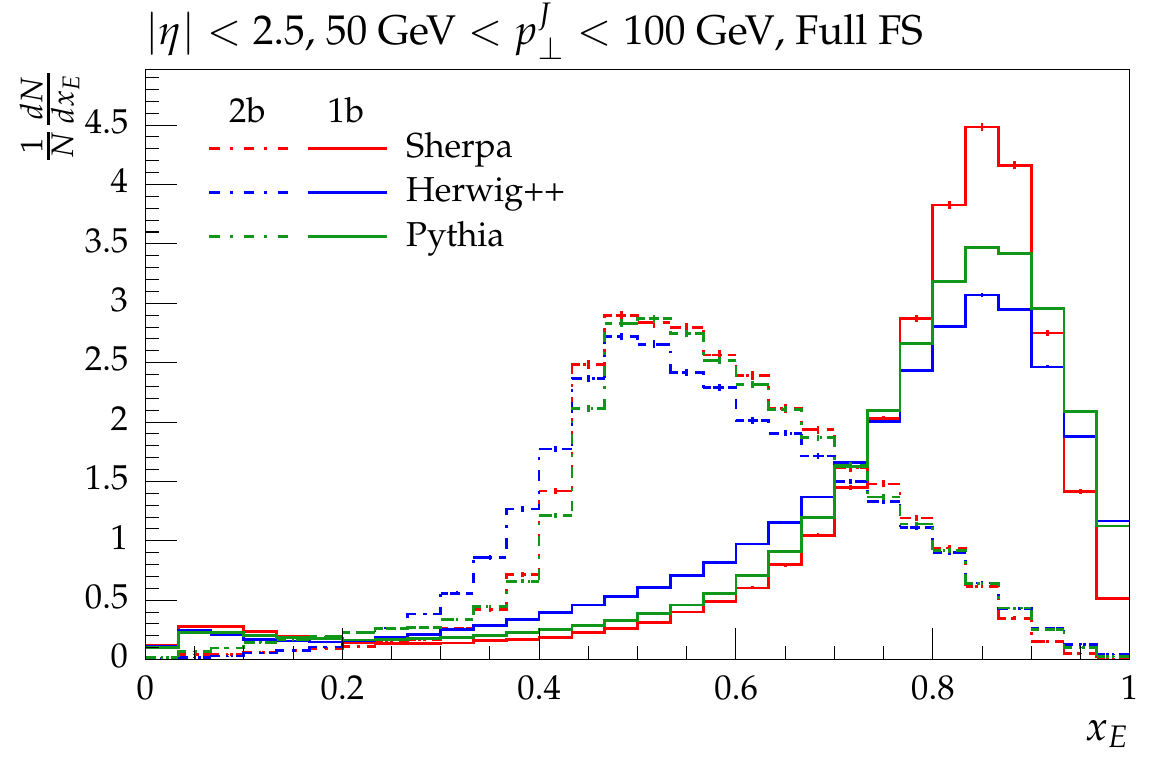}
  \includegraphics[scale=0.48]{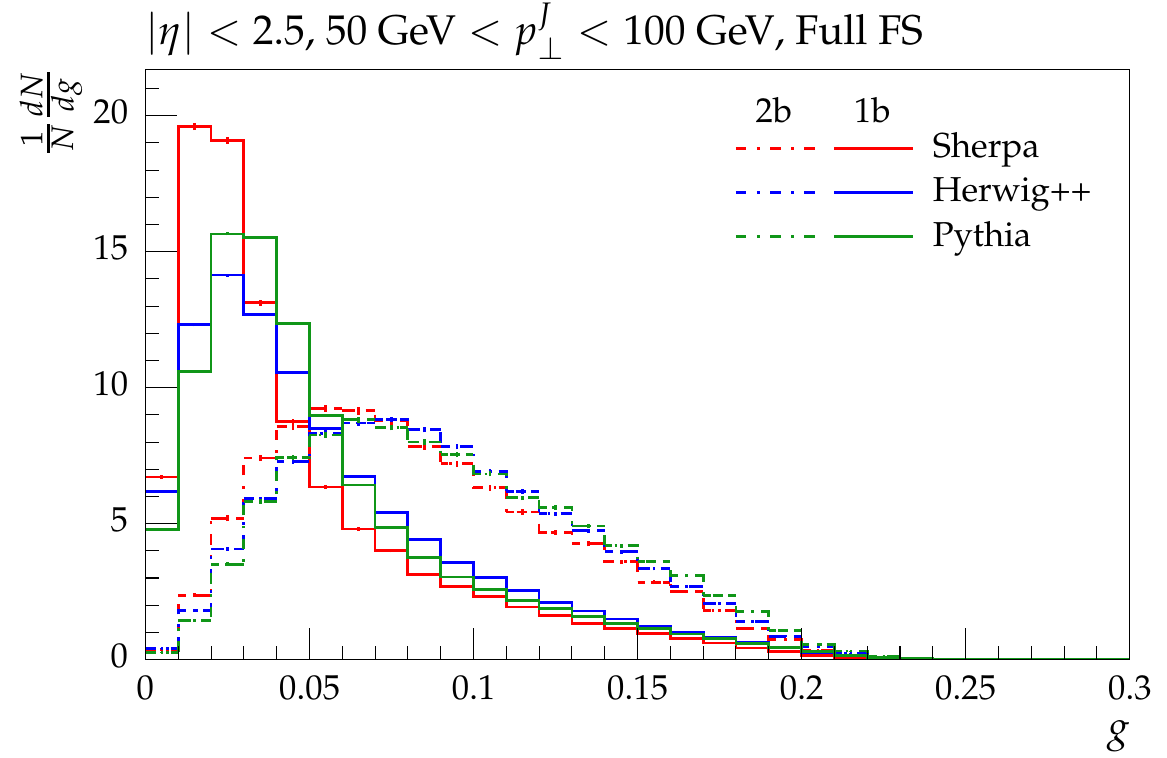}
  \includegraphics[scale=0.48]{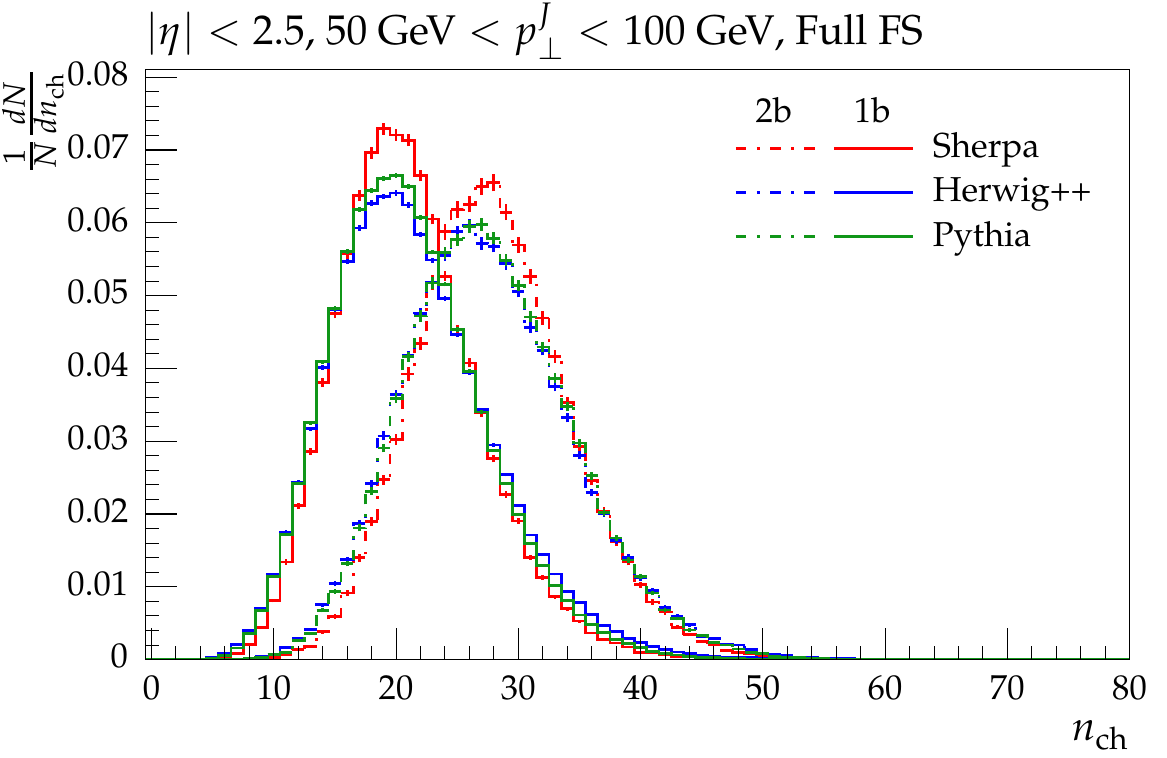}  
  \parbox{0.8\textwidth}{
    \caption{Comparison of $x_E$, $g$, $n_{{\rm ch}}$ from different event generators
      for jets within ${50\ \text{GeV} < p_{\perp}^{J} < 100}$~GeV, based on
      the full hadronic final state.  The left panel shows results for $x_E$ .
     The little enhancement at small values of $x_E$ stems from light jets,
      where a single $b$-quark was inserted, either through gluon splitting,
      where the other $b$-quark was radiated outside the jet, or by the
      underlying event. The central panel shows results for $g$, where the
      $b$-hadron(s) were set stable - this was only done in this plot
      and none of the rest of the paper.  The right panel shows the number
      of final state particles inside the jet (including uncharged ones).
      A vertex is defined as  having at least 3 tracks.
      }
  \label{fig:GenComparison}}
\end{figure}

A somewhat independent observable is related to the shape of the actual jet.
Based on the reasoning above, $b$-jets tend to be relatively narrow, with only
small amounts of radiation roughly following the direction of the colour
connection of the $b$-quark to the rest of the event.  In contrast
$b\bar{b}$-jets tend to originate from hard gluons, which may not only
radiate more due to the larger colour charge of $\CA=3$ vs.\ $\CF=4/3$
before they split, but which also have an intrinsic size related to the
relative distance of the two $b$-quarks inside the jet.  This effect could
be captured by using the mean of the energy distribution $\rho(r)$ inside
the jet, where $r<R$ is the radial distance of a hadron or similar to the
centroid of the jet with radius $R$.  It turns out, however, that a good
observable is provided by the first $p_\perp$-moment of this distribution
\begin{alignat}{5}
  g = \frac{1}{p_\perp^{(J)}}\,
  \sum\limits_{i \in\mbox{\rm Jet}} p_\perp^{(i)}\Delta R_{iJ}\;,
  \label{eq:girth}
\end{alignat}
an observable also known as ``girth'' $g$, or jet width.  Here $p_\perp^{(J)}$
is the transverse momentum of the jet, $p_\perp^{(i)}$ the transverse momentum
of the hadron, track, or energy cell ${(i)}$ inside the jet
$(i\in\mbox{\rm Jet})$, and $\Delta R_{iJ}$ is its radial distance with respect
to the jet vector. 

Many more observables can be used with different distinguishing powers and
robustness. A prime example is the number of charged tracks $n_{{\rm ch}}$.
Despite presenting a possibly poor Monte Carlo modelling, highly depending on
the details of hadronization modelling and underlying event implementation,
they are still  extensively used by experimental analyses.  Hence, we also
inspect its impact in the following section. 

The typical behaviour of these observables is exemplified in
Fig.~\ref{fig:GenComparison}; in this figure all jets have a transverse
momentum $p_\perp^J$ between $50$ and $100\,\UGeV$ and their pseudorapidity
$|\eta_J|<2.5$. To provide an idea of modelling uncertainties, the results of
different event generators, \HerwigPP~\cite{Bahr:2008pv},
\PythiaEight~\cite{Sjostrand:2014zea} and \Sherpa~\cite{Gleisberg:2008ta}
are exhibited.

There are other observables that aim to scrutinize the colour connection and
2-dimensional shape of the jet, \eg, planar flow, pull or differential jet
shape that were also inspected.  However, in this study only the most powerful
observables will be investigated, namely fragmentation fractions $x_E$,
girth $g$ and number of charged tracks $n_{{\rm ch}}$.  These additional
observables could be used in the construction of more advanced discriminators
based on boosted decision trees or neural networks, which is beyond the scope
of this study.  It is worth noting that there are interesting similarities
between the investigations here and studies aiming at distinguishing gluon
and light quark jets, see for instance Ref.~\cite{Gallicchio:2011xq,
  Gallicchio:2012ez}.   However, for obvious reasons, in their case the
fragmentation fraction does not result in sizable improvements to the
efficiencies for gluon vs.\ light quark tagging.

\section{Analysis}
\label{sec:analysis}

\begin{figure}[b!]
\includegraphics[scale=0.43]{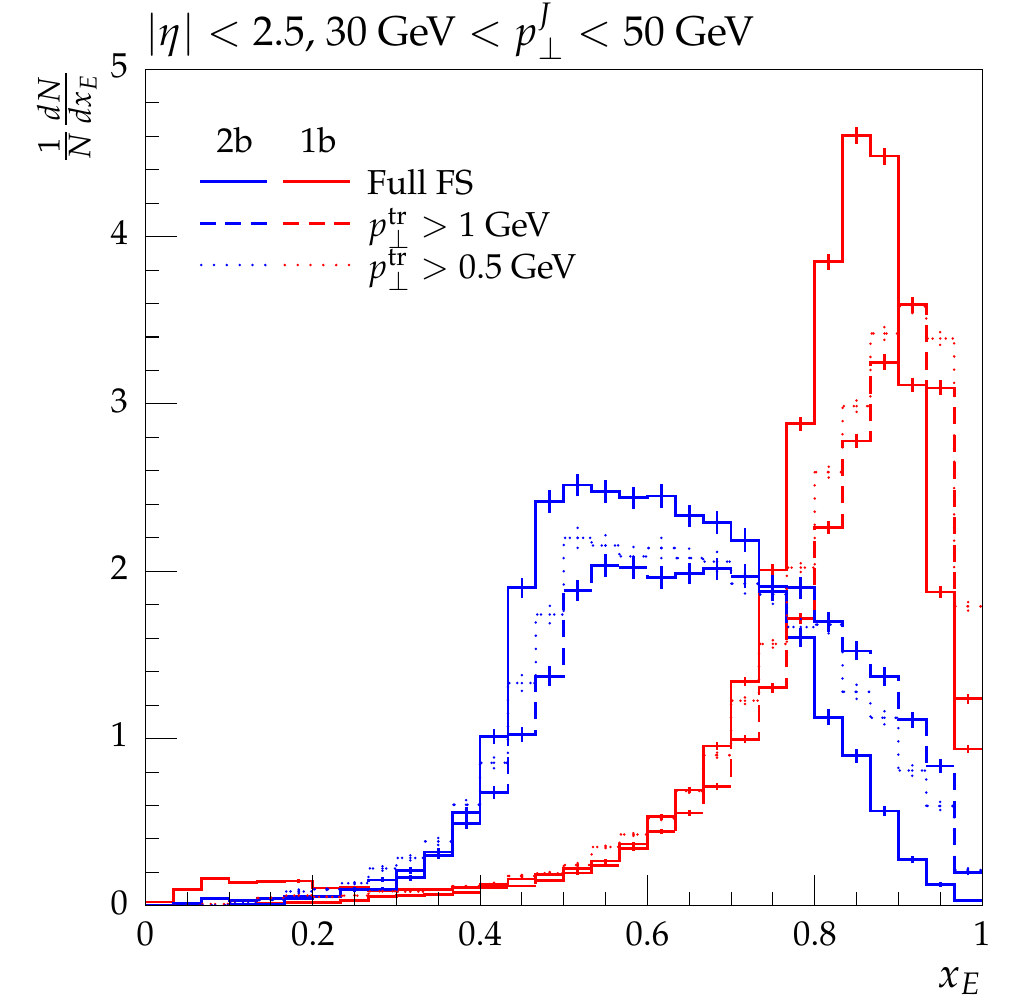}\hspace{-0.3cm}
\includegraphics[scale=0.43]{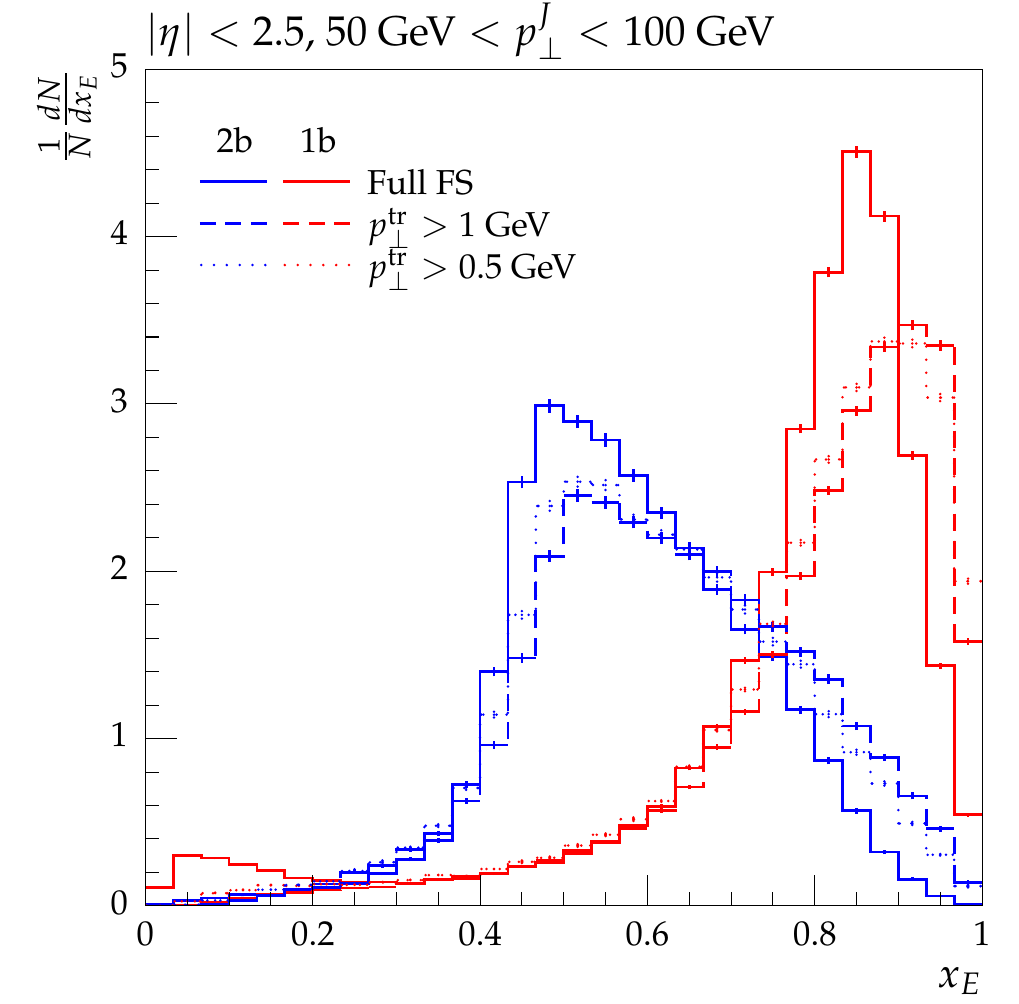}\hspace{-0.3cm}
\includegraphics[scale=0.43]{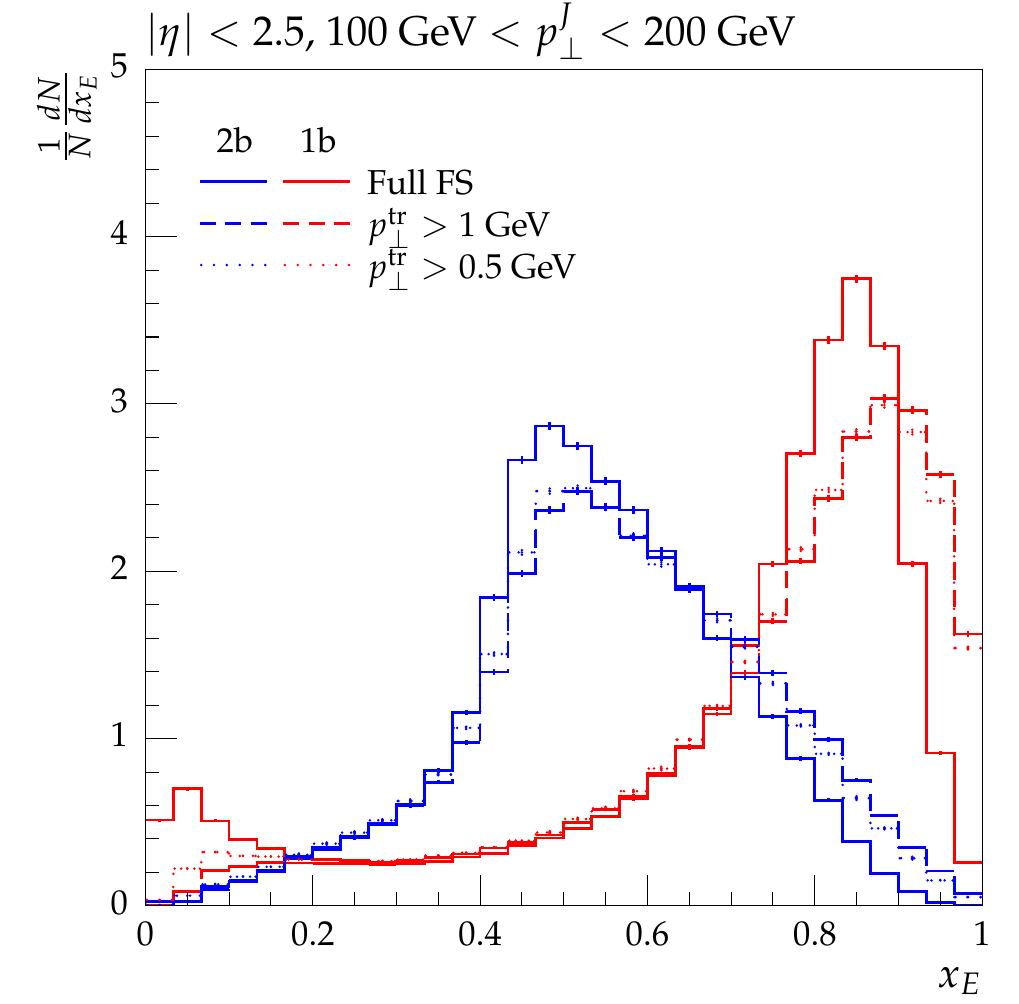}\hspace{-0.3cm}
\includegraphics[scale=0.43]{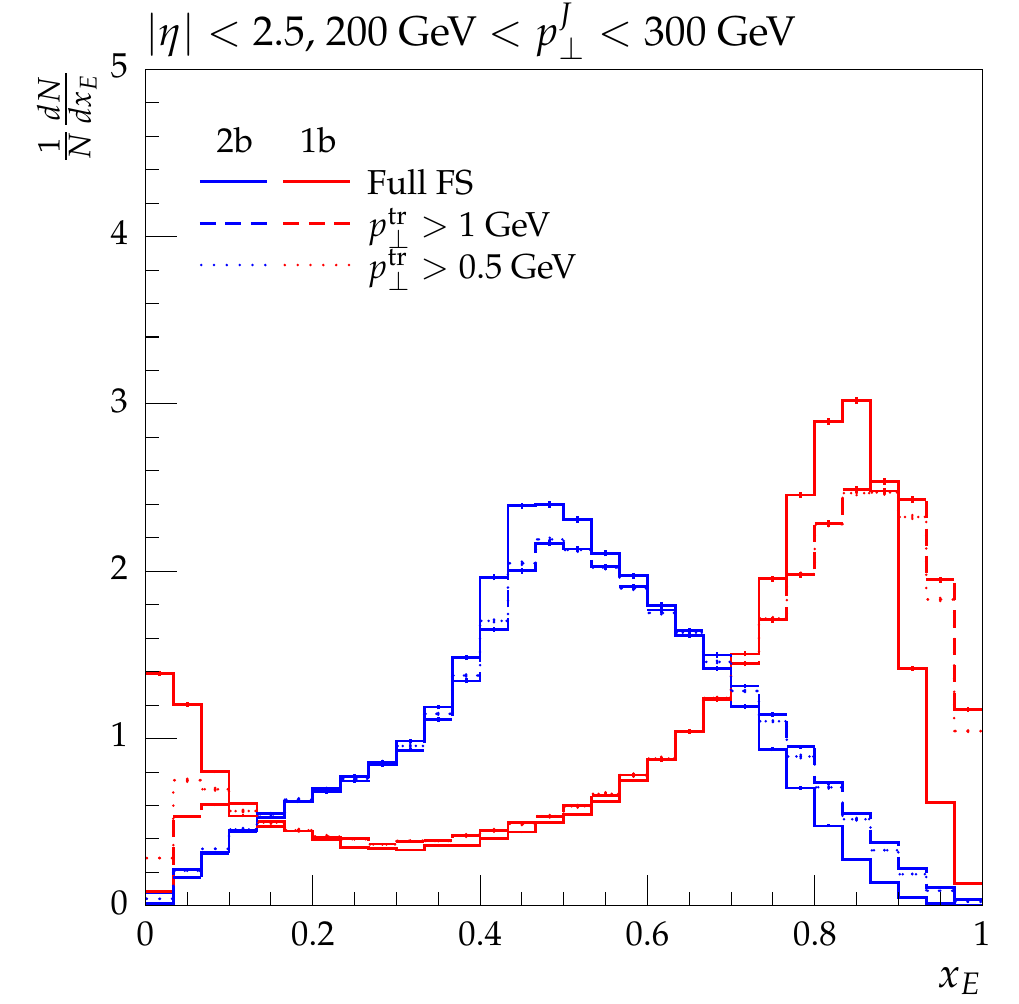}\hspace{-0.3cm}
\includegraphics[scale=0.43]{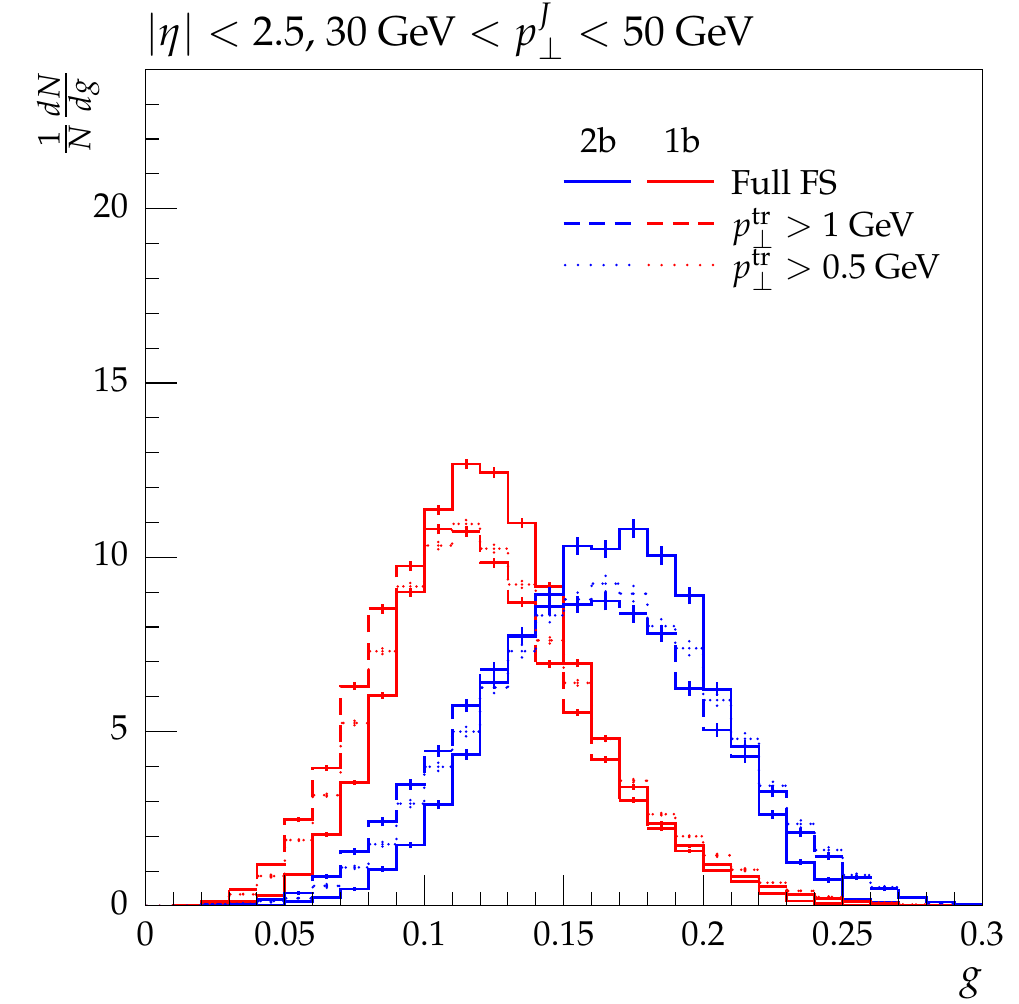}\hspace{-0.3cm}
\includegraphics[scale=0.43]{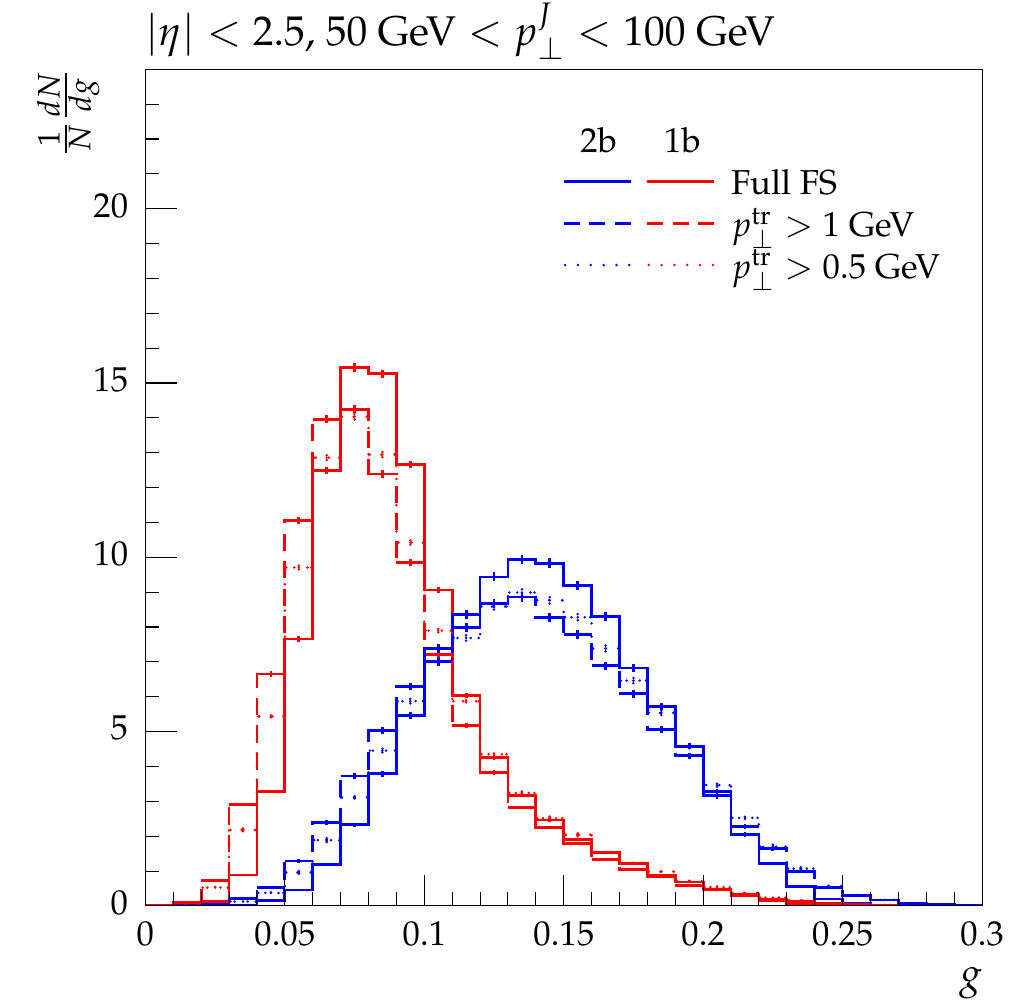}\hspace{-0.3cm}
\includegraphics[scale=0.43]{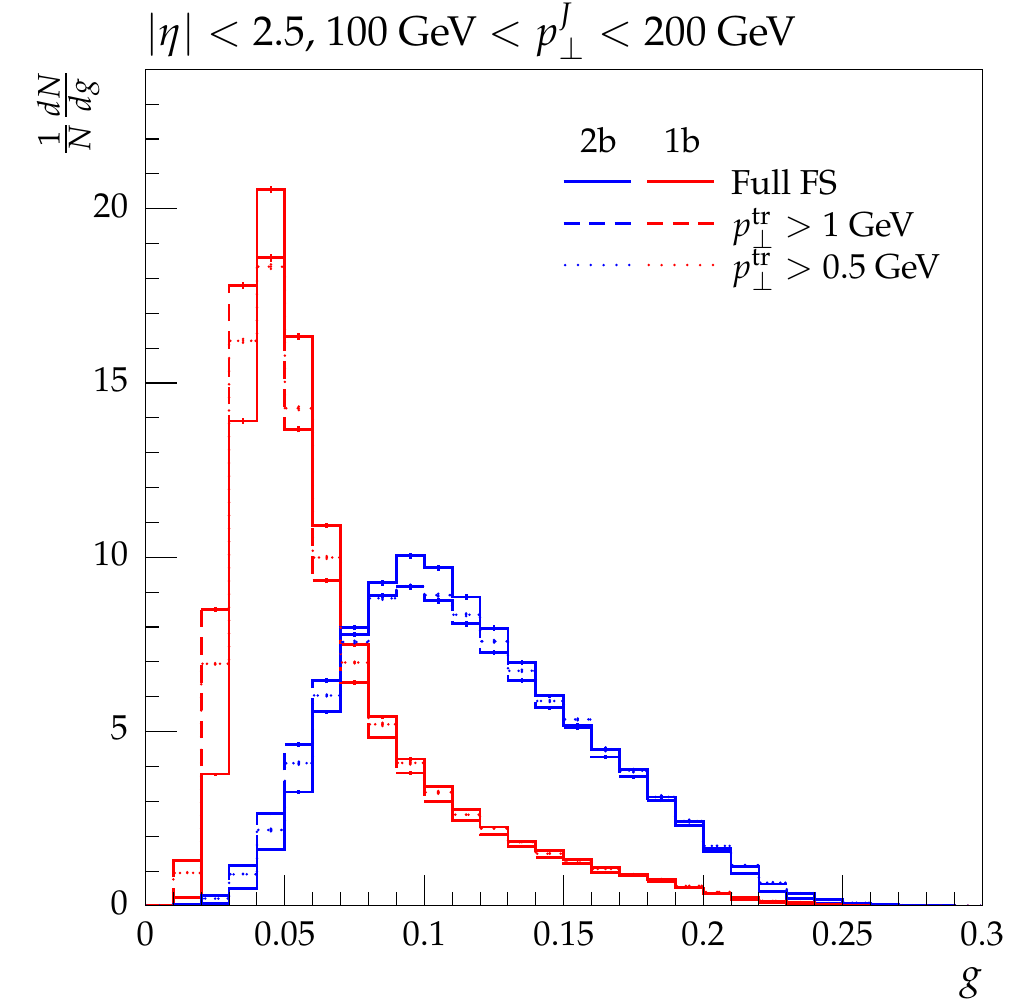}\hspace{-0.3cm}
\includegraphics[scale=0.43]{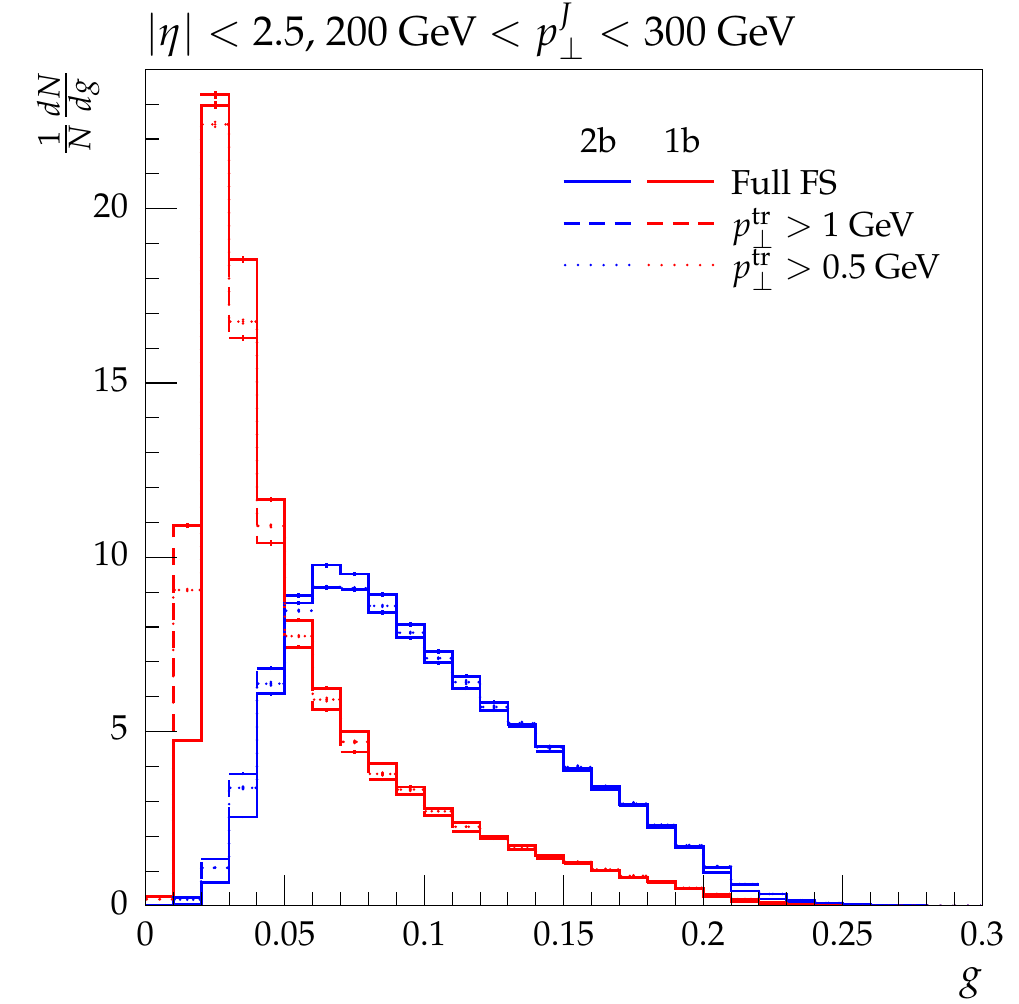}\hspace{-0.3cm}
\includegraphics[scale=0.43]{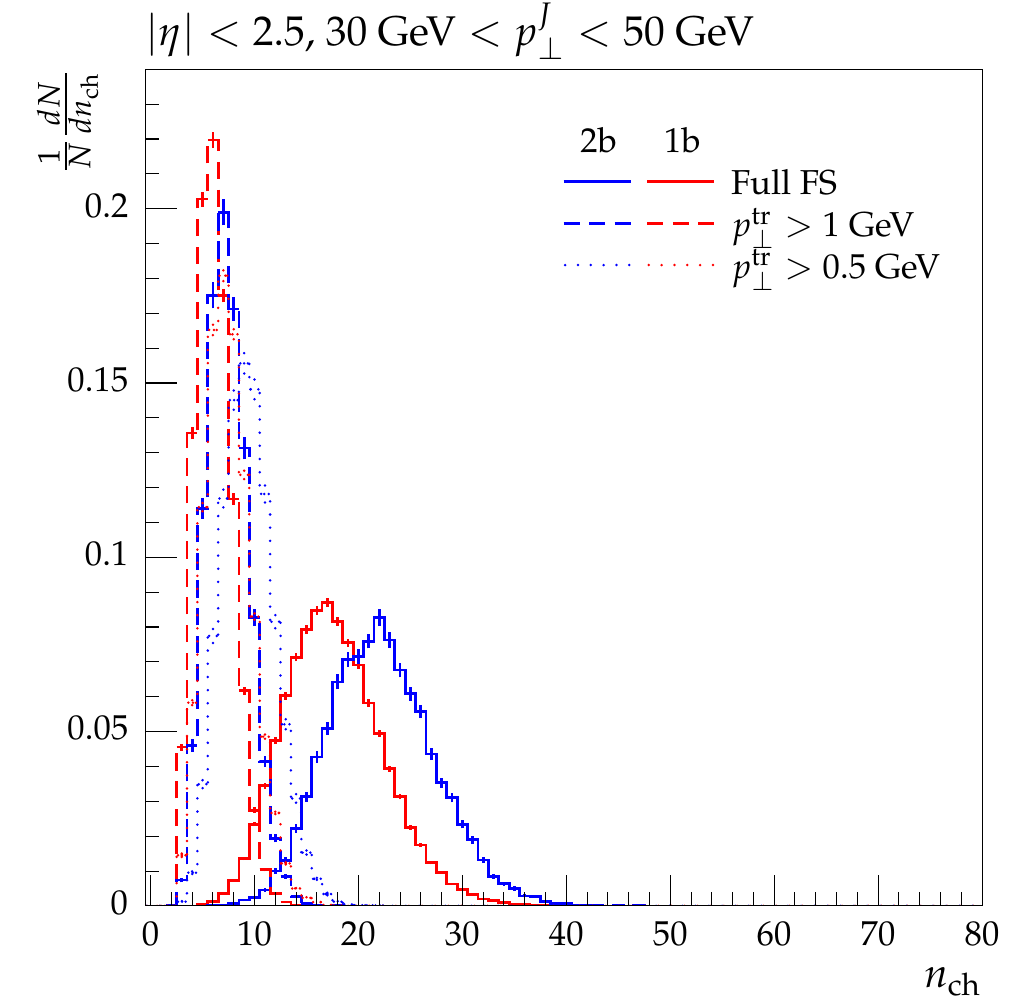}\hspace{-0.3cm}
\includegraphics[scale=0.43]{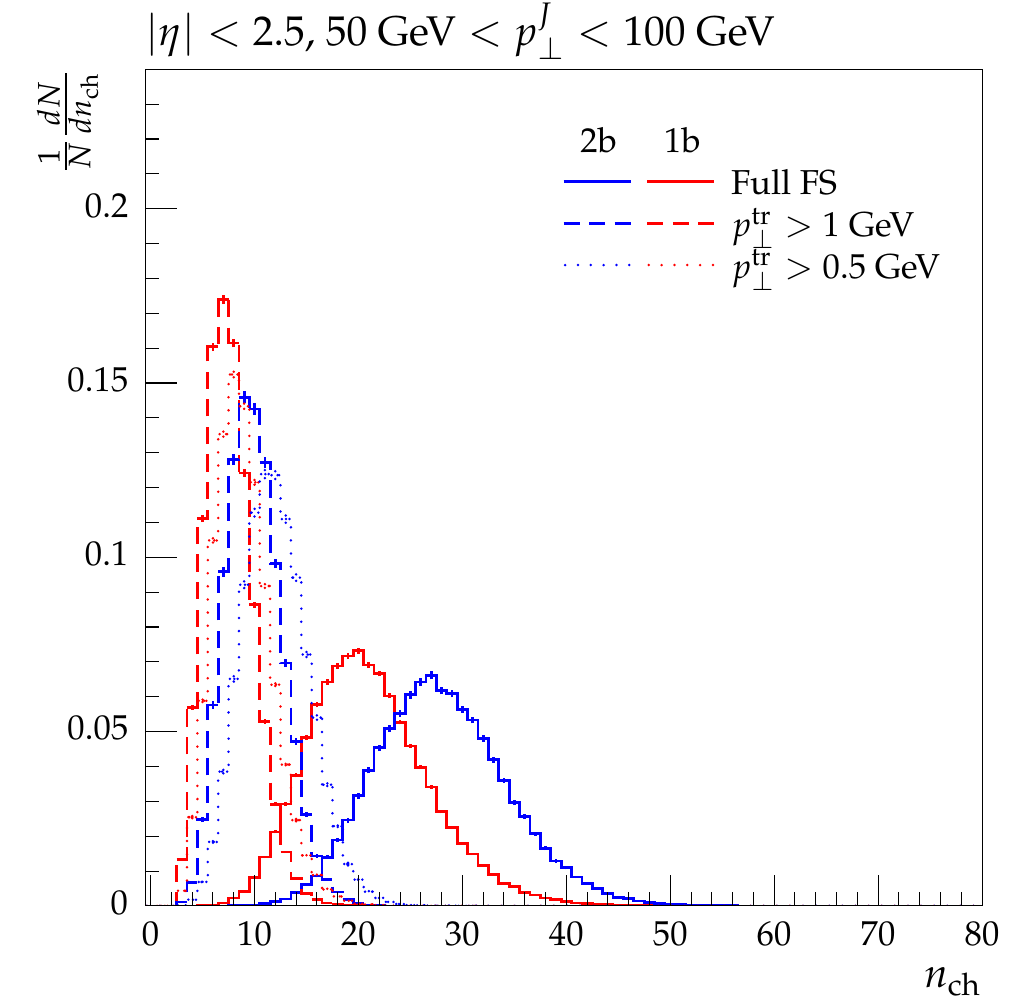}\hspace{-0.3cm}
\includegraphics[scale=0.43]{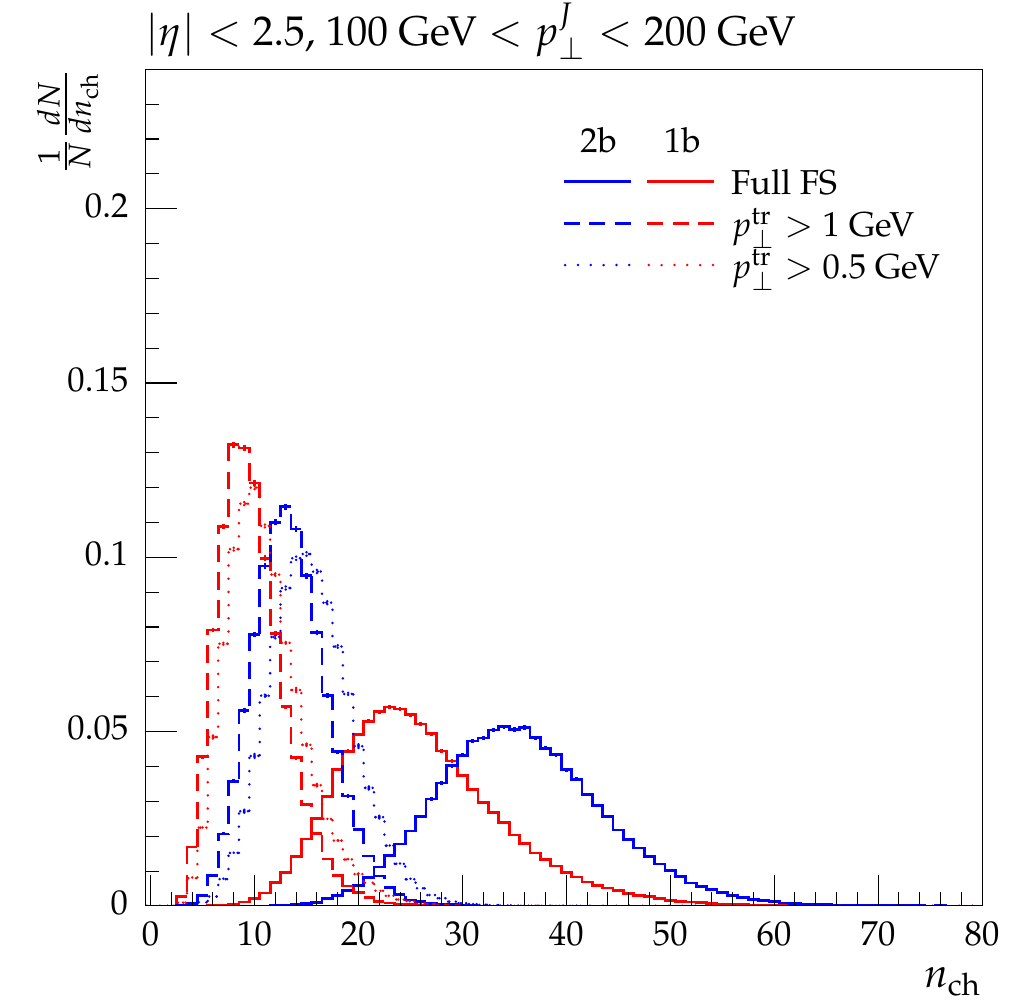}\hspace{-0.3cm}
\includegraphics[scale=0.43]{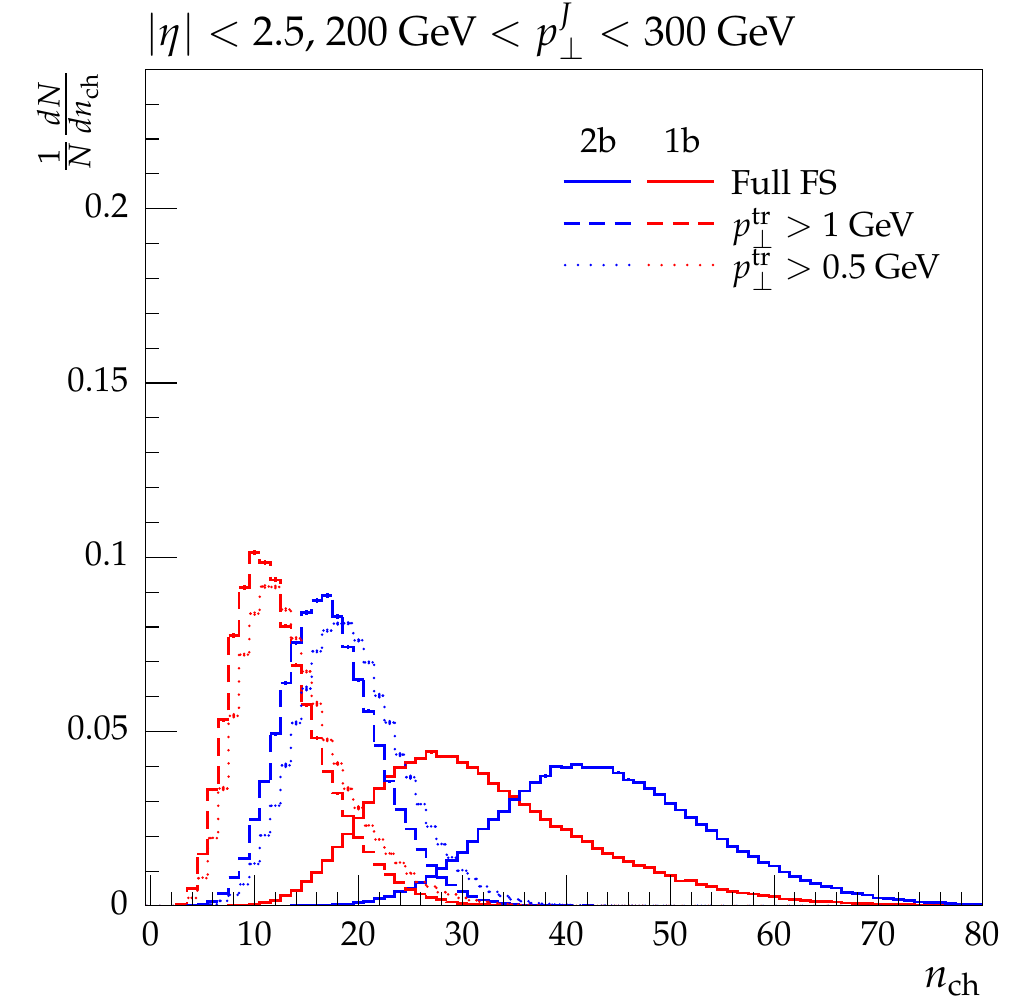}\hspace{-0.3cm}
  \parbox{0.8\textwidth}{
\caption{$x_E$ (top row), girth (central row) and number of charged tracks (bottom row) distributions 
for jets within different $p_\perp$ slices: ${30\ \text{GeV} < p_{\perp}^{J} < 50}$~GeV (first column) ${50\ \text{GeV}
 < p_{\perp}^{J} < 100}$~GeV (second column), ${100\ \text{GeV} < p_{\perp}^{J} < 200}$~GeV (third column) and ${200\
 \text{GeV} < p_{\perp}^{J} < 300}$~GeV (fourth column). Red curves correspond to jets with one $b$-hadron and blue with
 two $b$-hadrons. Solid lines are based on the full hadronic final state, including uncharged  particles, dashed
 lines on charged tracks with a minimum $p_\perp^{{\rm tr}}$ of 1 GeV and dotted with a minimum $p_\perp^{{\rm tr}}$ of 0.5~GeV.
 A vertex is defined as  having at least 3 tracks.}
\label{fig:EFragmentation1}}
\end{figure}

As a test case, a pure QCD $pp\rightarrow$~jets sample at the
$\sqrt{s}=13$~TeV LHC is considered.  The event sample was generated with 
\Sherpa~\cite{Gleisberg:2008ta} in a very basic setup, using $2\rightarrow 2$
matrix elements at leading order, supplemented with the default parton shower
based on Catani--\-Seymour subtraction~\cite{Schumann:2007mg}, and accounting
for hadronization and underlying event effects.  Since different event
generators differ in their approximations and implementation details of the
parton shower evolution and non-perturbative models it is important to
quantify the resulting uncertainties and to access the robustness of the
results.  To this end, event samples with the same specifications have been
generated and analysed, using \HerwigPP~\cite{Bahr:2008pv} and
\PythiaEight~\cite{Sjostrand:2014zea}.  Where relevant, the results from
these different simulation tools are contrasted; overall, however, they do
not impact on the results and conclusions of this study.

The analysis is performed using~\Rivet~\cite{Buckley:2010}.  Jets are defined
by the anti-$k_T$ algorithm, using \Fastjet~\cite{Cacciari:2008gp}, with
$R = 0.4$, requiring $p_{\perp}^J>30$~GeV and $|\eta_J|<2.5$.  Charged tracks
are defined with a minimum transverse momentum
$p_\perp^{{\rm tr}}\ge 0.5$ or $1\,\UGeV$.  The different cutoffs are used to
probe the stability of the observables.  Lowering the threshold would of course
lead to more statistics, however, it also increases the dependence on the MC
modelling.

Jets are categorized as containing one or two $b$-hadrons, with other values
rejected, counting their number inside the jet radius.  For our purposes,
the $b$-hadrons are ``reconstructed'' from the event record, taking into
account the choice of observable final state particles.  In case of two
different $b$-hadrons in the jet, by default the harder one is selected.  
\begin{figure}[b!]
\includegraphics[scale=0.48]{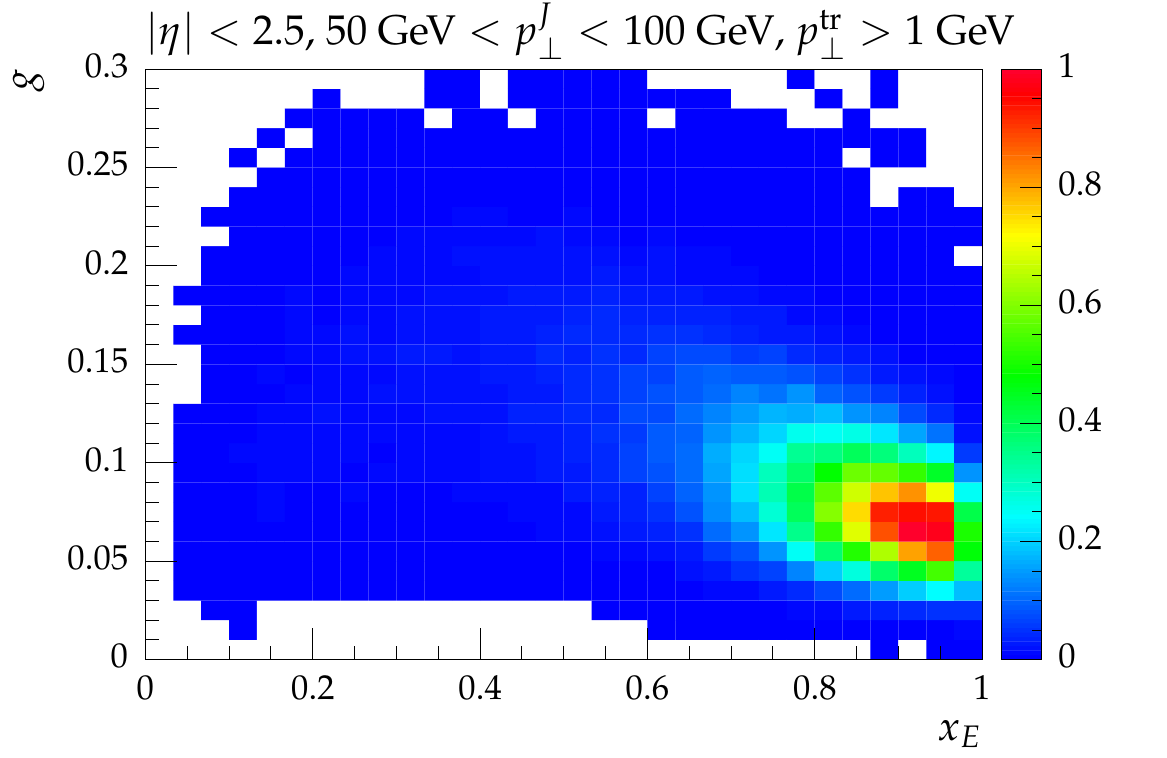}
\includegraphics[scale=0.48]{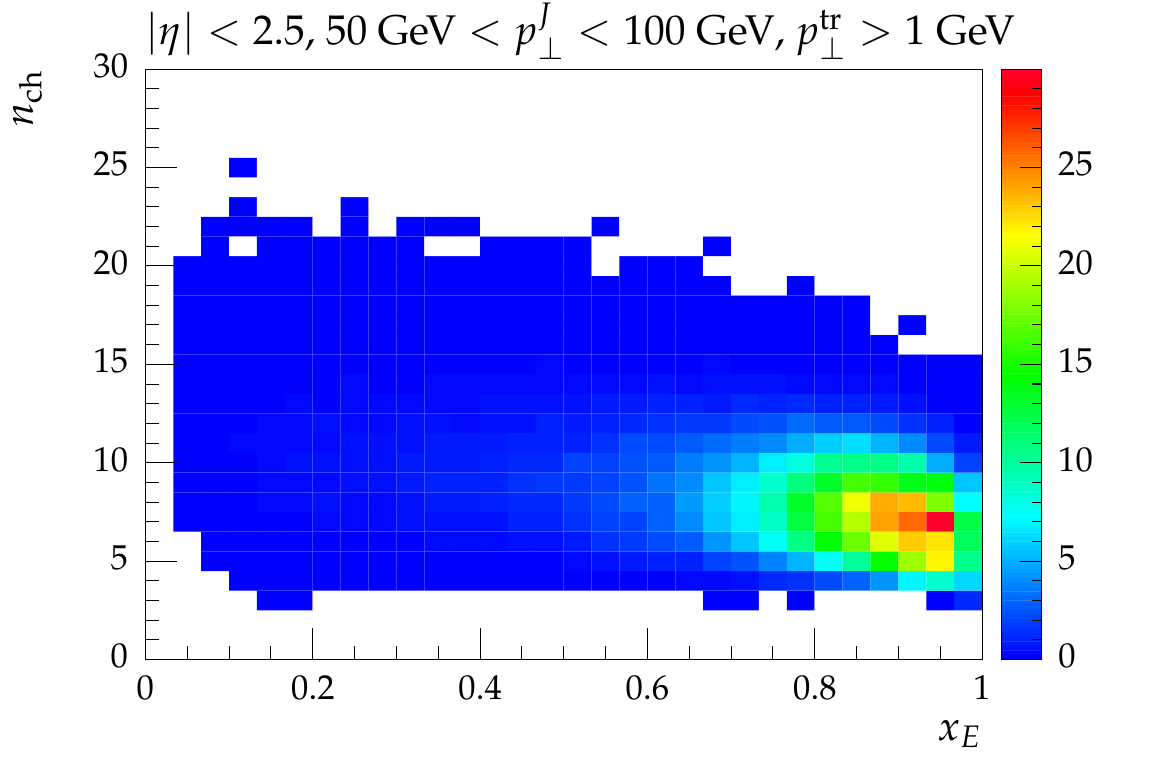}
\includegraphics[scale=0.48]{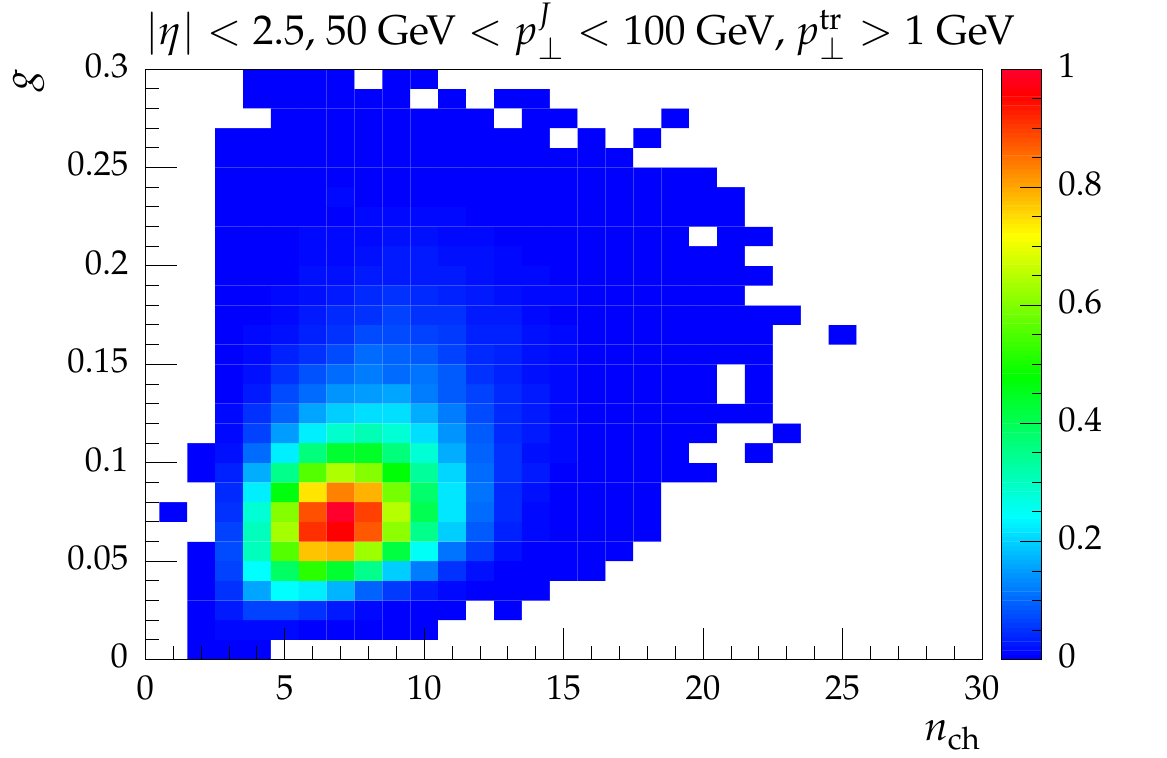}
\includegraphics[scale=0.48]{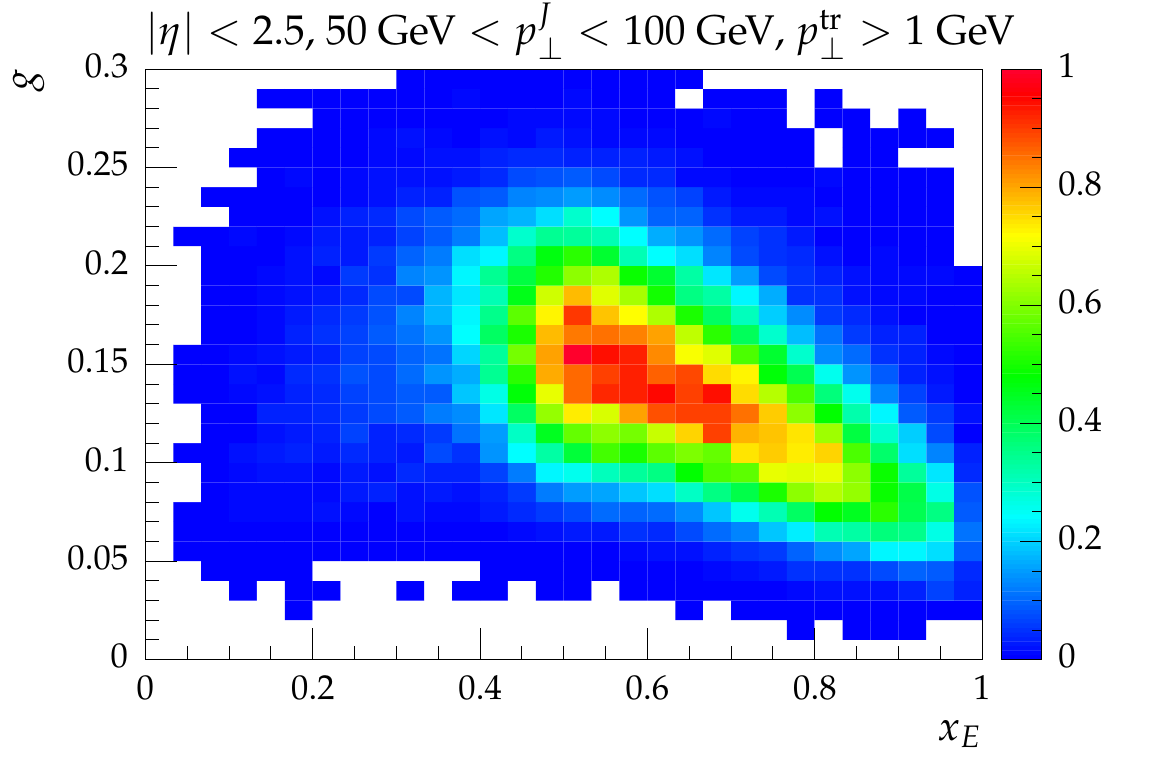}
\includegraphics[scale=0.48]{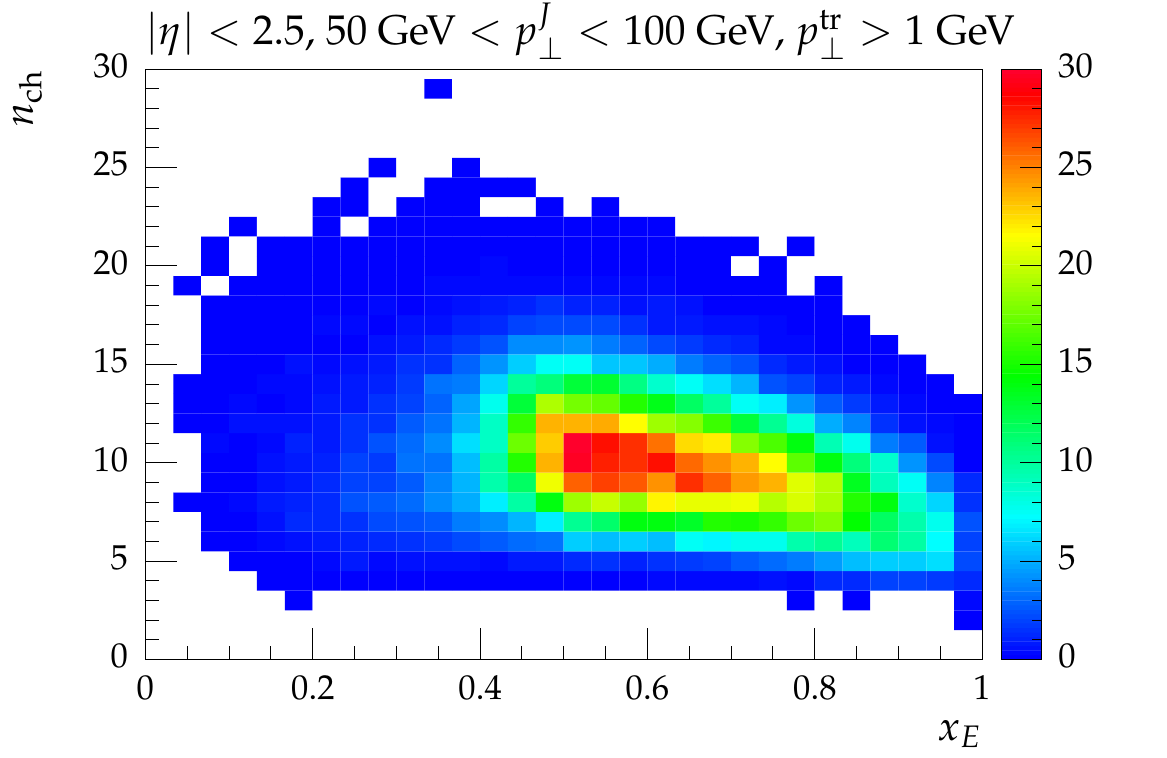}
\includegraphics[scale=0.48]{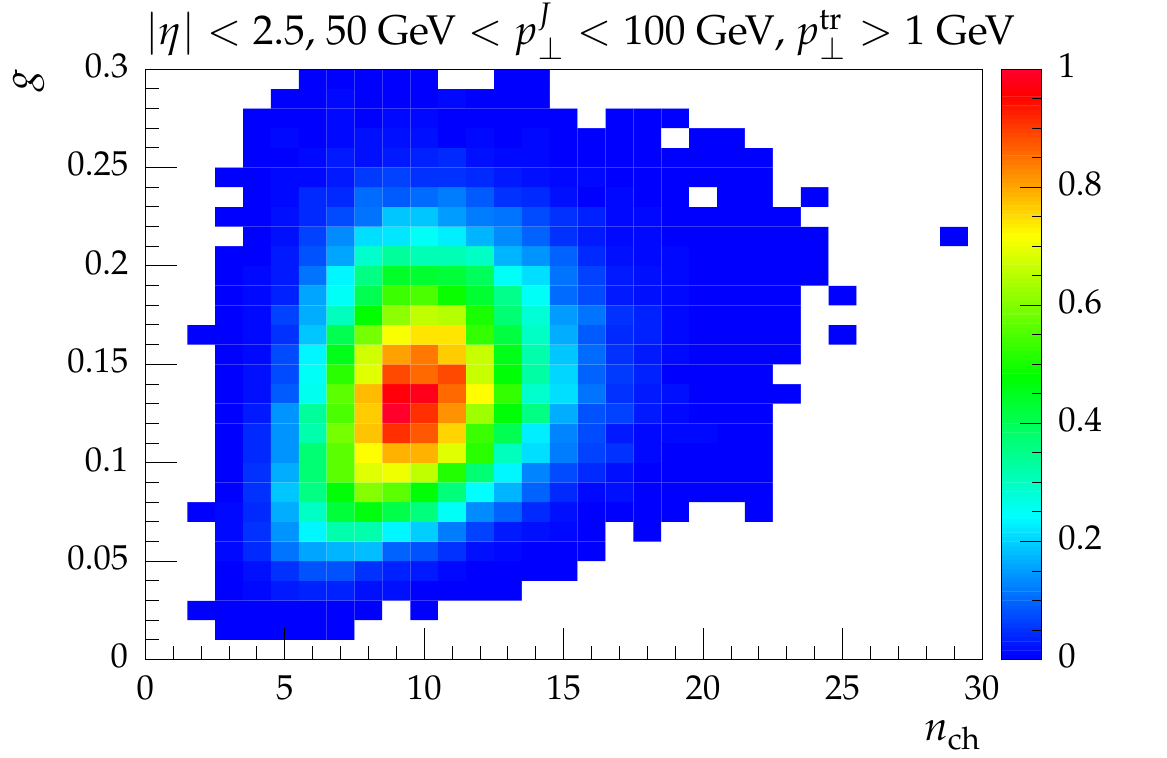}
\parbox{0.8\textwidth}{
  \caption{Correlations between the fragmentation fraction with the girth
    $(x_E,\,g)$ (left column),  fragmentation fraction with charged tracks
    $(x_E,n_{{\rm ch}})$ (central column) and charged tracks with girth
    $(g,n_{{\rm ch}})$ (right column).  The colours represent the normalized
    weight of the particular bin.  The top plots are for one $b$-hadron in
    the jet, the bottom ones for two $b$-hadrons in the jet.  The jets
    considered here have a $p_\perp^J$ of 50 to 100 GeV. The objects considered
    in the analysis in this case are charged tracks of at least 1~GeV
    $p_\perp^{{\rm tr}}$. A vertex is defined as having at least 3 tracks.}
  \label{fig:Correlation}}
\end{figure}

In Fig. \ref{fig:EFragmentation1} (top row), the $x_E$ distributions are displayed. It is observed that in the case of one $b$-hadron in the jet,
the $b$-hadron carries most of the energy content with the distribution peaking between 0.8 and 1, depending on the $p_\perp$ slice. 
On the other hand, in the case of two $b$-hadrons in the jet  the energy fraction for the most energetic $b$-hadron tend to be near 0.5 -- 0.6. 
These effects do not diminish when considering  only charged tracks, rather it improves slightly, 
\eg, the distribution for one $b$-quark in the jet narrows near $x_E=1$. Similar observables built out of the 3-momentum, 
the transverse momentum or weighted with the cosine of the angle to the jet axis present qualitatively and quantitatively 
similarities to $x_E$. Therefore, only the latter is considered for simplicity.

The girth distributions $g$ are displayed in Fig.~\ref{fig:EFragmentation1} (central row). This observable presents a good
separation between the single and double b-tagging case.  The double b-tag sample leads to broader jets
in respect to the single b-tag case.  This observable presents useful results at either low or high $p_\perp^J$. 
Moreover, the charged tracks present qualitatively similar results and only a subleading dependence on the 
threshold energy, ${p_\perp^{{\rm tr}}>0.5}$~GeV or 1~GeV, is observed.

The dependences on the charged track multiplicity $n_{{\rm ch}}$ is inspected in Fig.~\ref{fig:EFragmentation1} (bottom row). 
The jets with two $b$-hadrons present a much higher multiplicity than the single b-tagged. This is a result of the longer
decay chain of the $b$-hadron and the different emission pattern described by the parton shower. These differences are 
enhanced at higher $p_\perp^J$ where the $n_{{\rm ch}}^{2b}/n_{{\rm ch}}^{1b}$ slowly converges to $\CA/\CF$. 

Despite $n_{{\rm ch}}$ not being an infrared safe observable and therefore highly dependent on the parton shower, hadronization
and underlying event modelling, the disagreement with the MCs is usually suppressed via an appropriate tuning to the LHC data. 
Hence, its applications  have to account for these limitations and/or should be taken with a grain of salt.


\section{Double and Single b-tagging Efficiencies}
\label{sec:combination}

\begin{figure}[b!]
\includegraphics[scale=0.43]{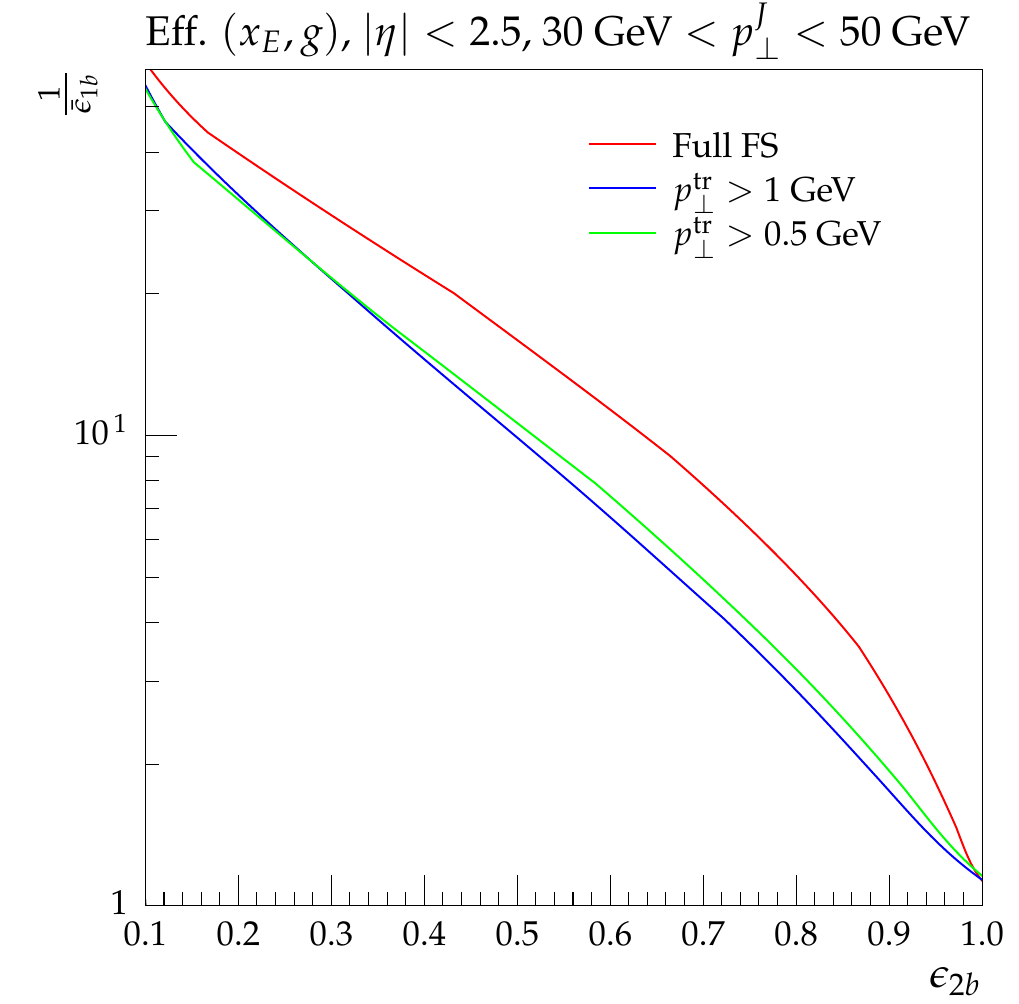} \hspace{-0.3cm}
\includegraphics[scale=0.43]{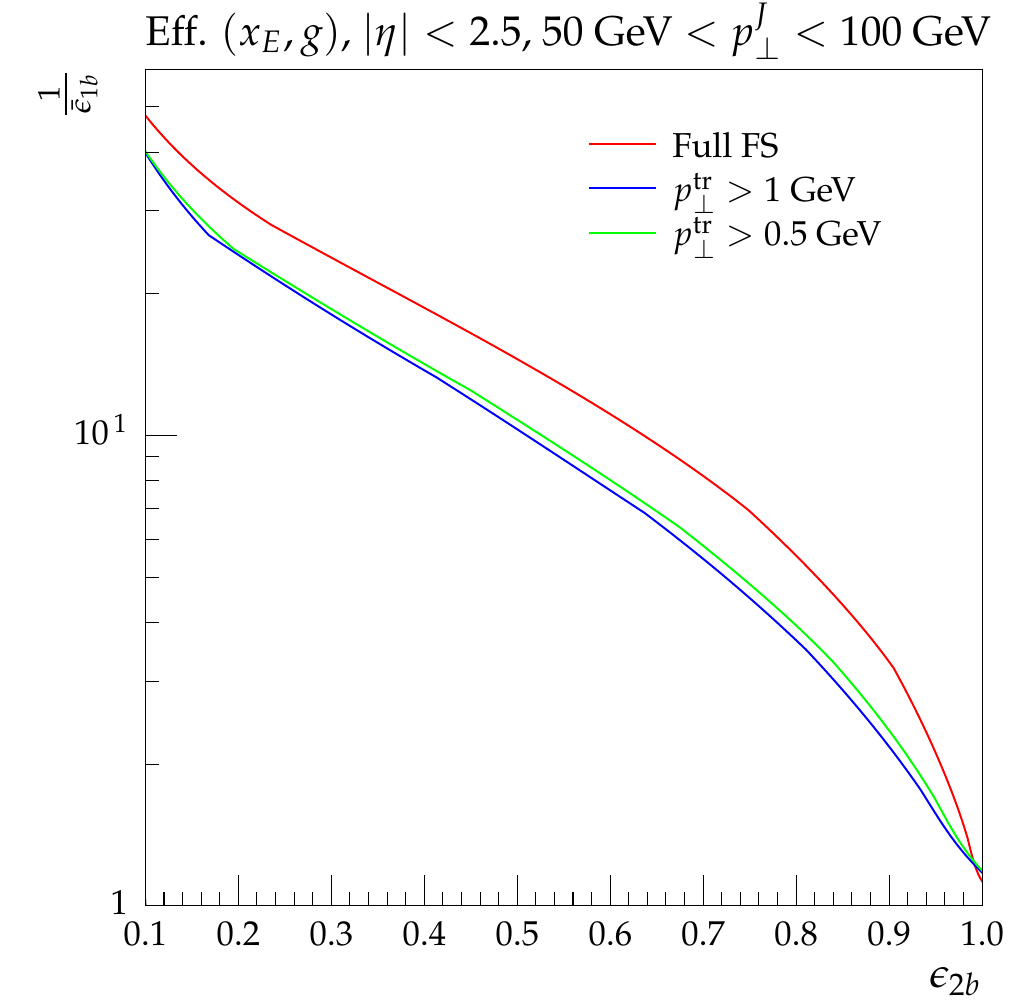}\hspace{-0.3cm}
\includegraphics[scale=0.43]{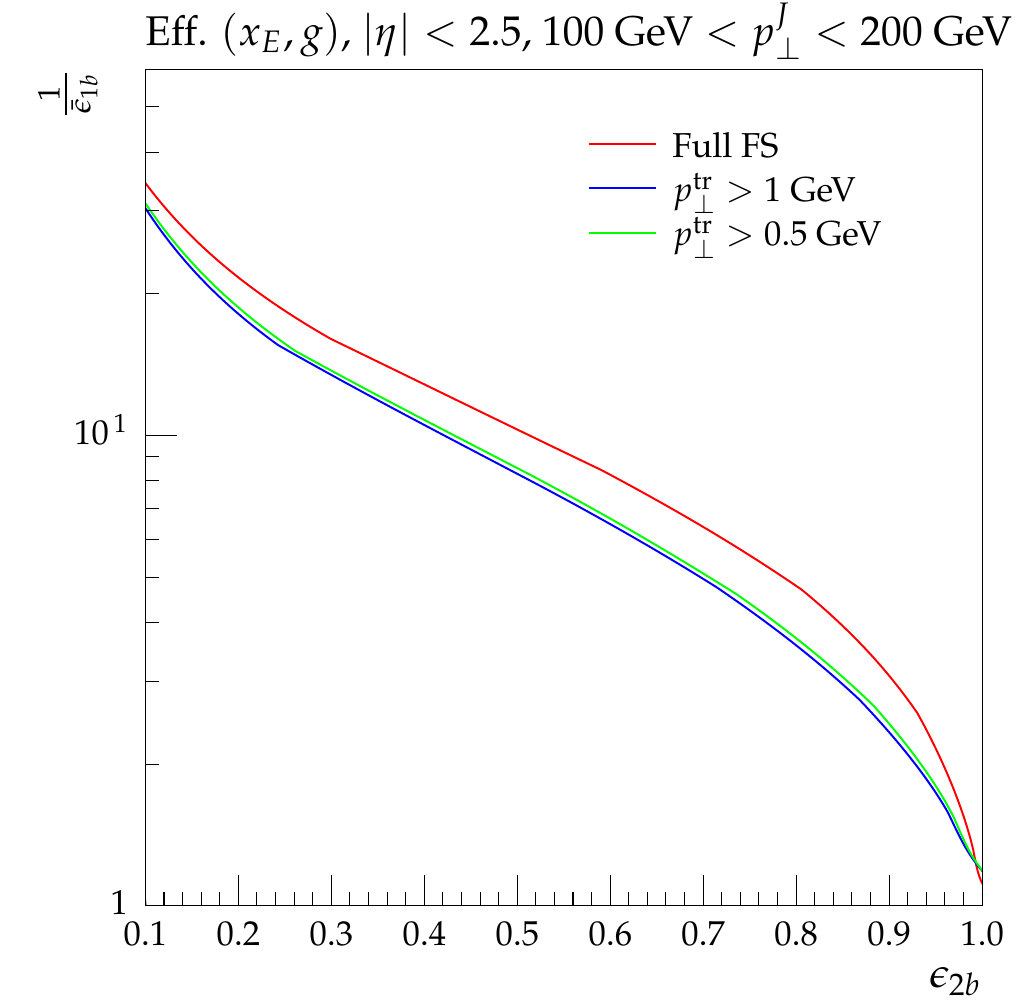}\hspace{-0.3cm}
\includegraphics[scale=0.43]{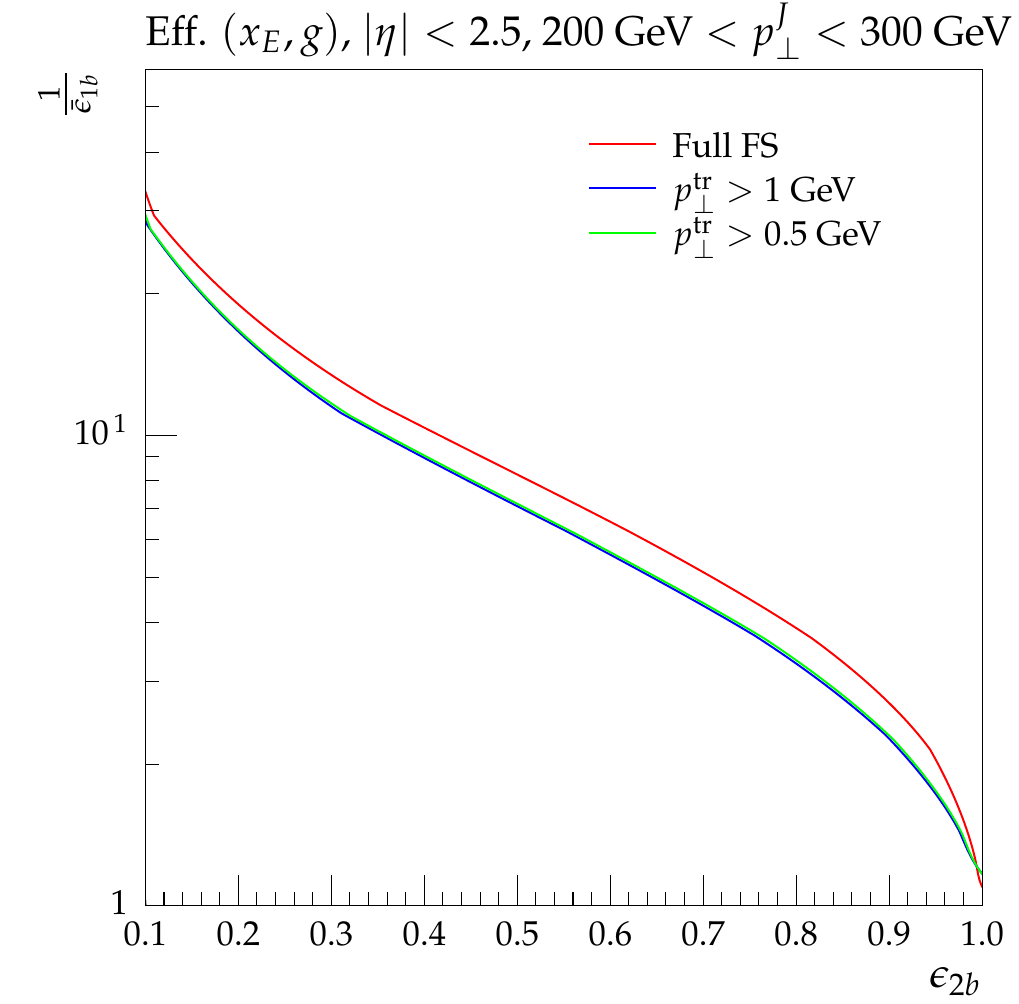}\hspace{-0.3cm}
\includegraphics[scale=0.43]{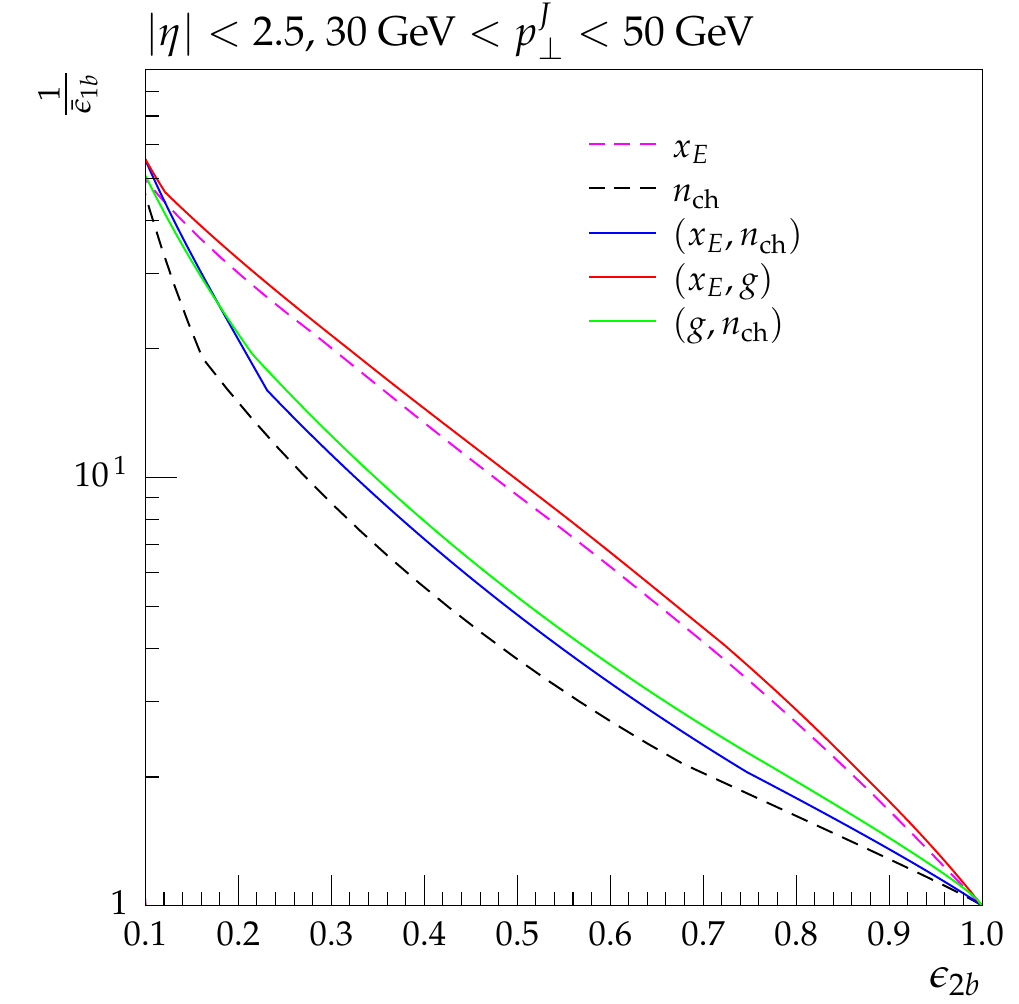}\hspace{-0.3cm}
\includegraphics[scale=0.43]{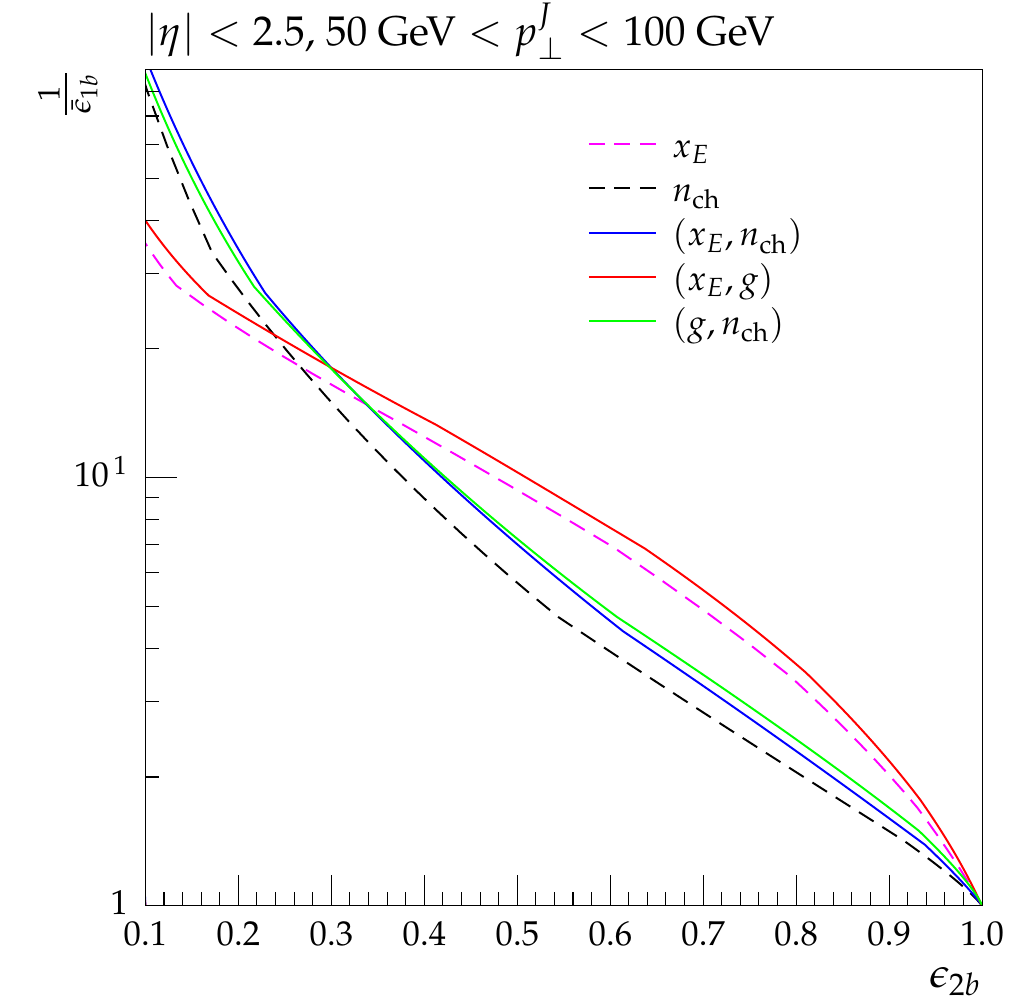}\hspace{-0.3cm}
\includegraphics[scale=0.43]{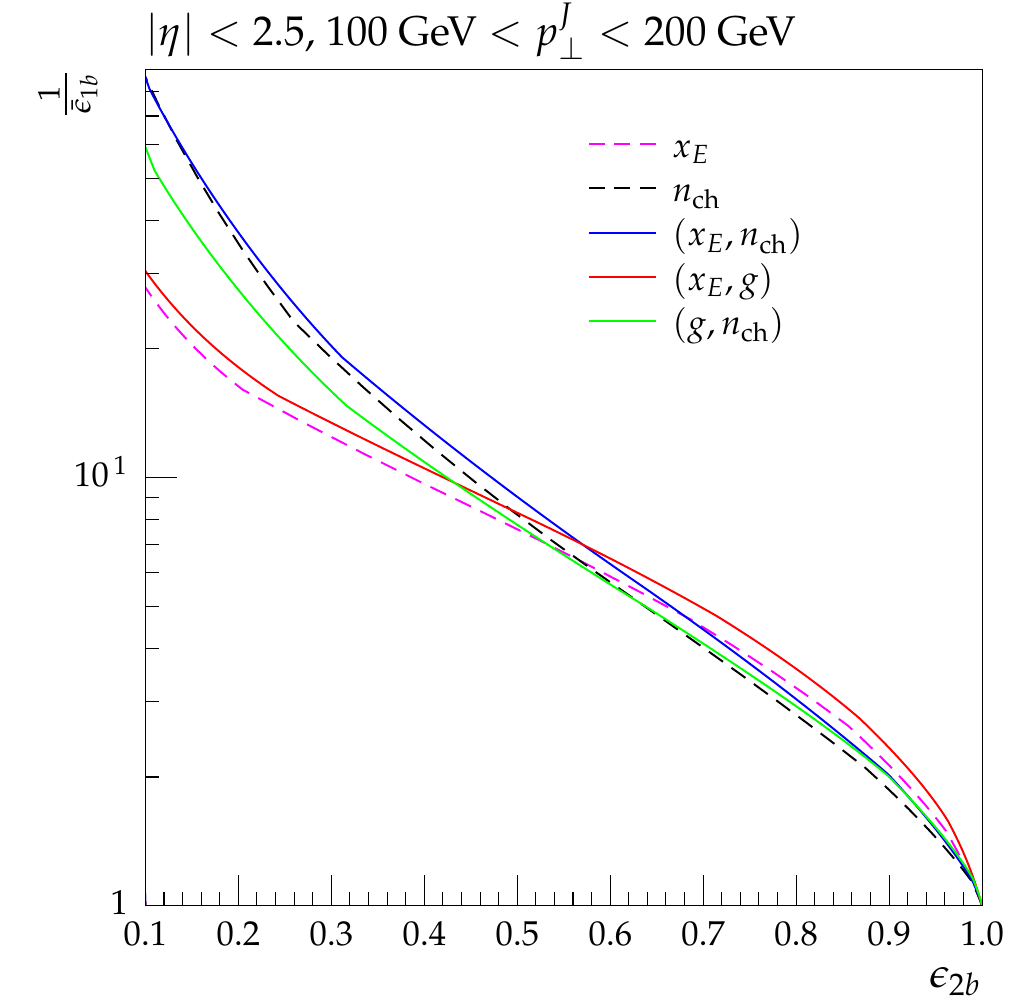}\hspace{-0.3cm}
\includegraphics[scale=0.43]{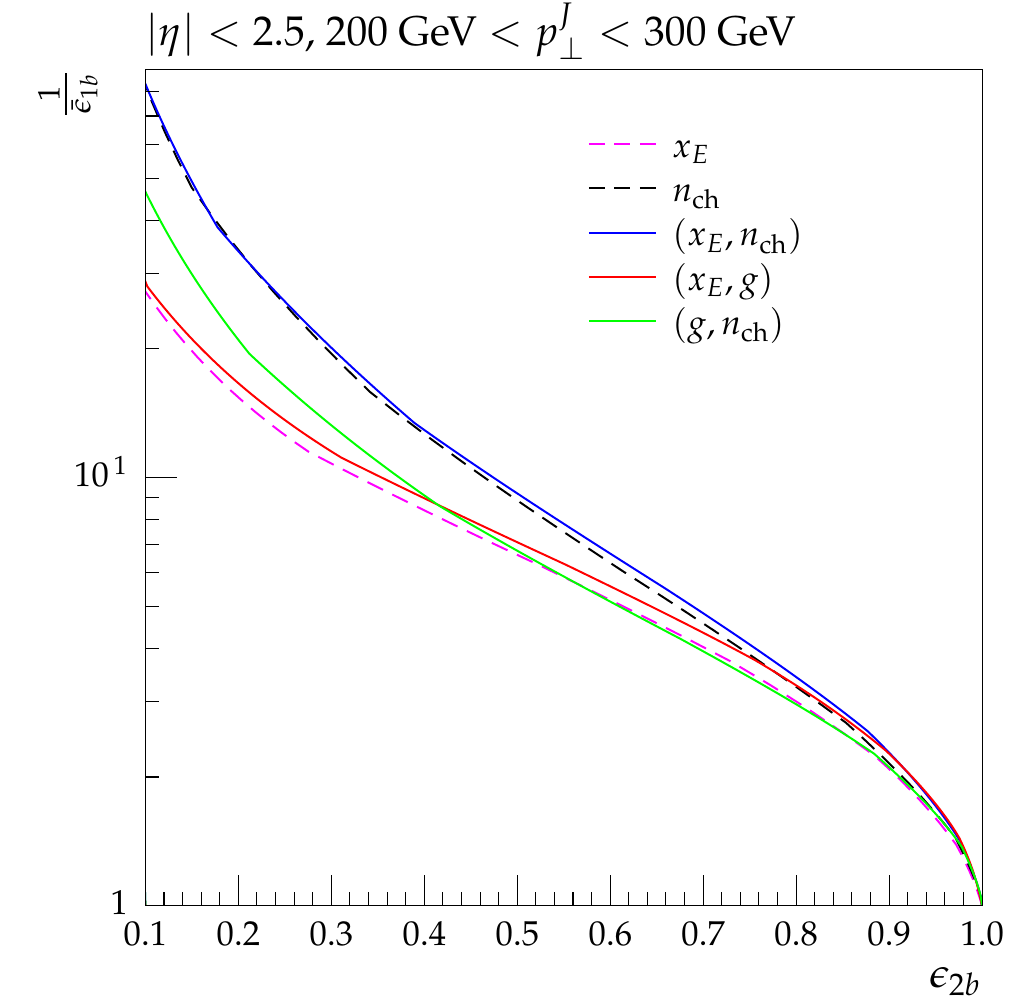}\hspace{-0.3cm}
  \parbox{0.8\textwidth}{
    \caption{Efficiency for tagging a $b$-jet as containing two $b$-hadrons
      $\epsilon_{2b}$ against the rejection of jets containing one $b$-hadron
      $1/\bar{\epsilon}_{1b}$ from combining $x_E$ and girth. The plots are again
      shown in different $p_\perp^J$ bins as in Fig.~\ref{fig:EFragmentation1}.
      Top row: The red curves refer to an analysis using the full final state,
      whereas the blue and green consider only charged tracks with minimum
      $p_\perp^{{\rm tr}}$ of 1~GeV and 0.5~GeV, respectively. Bottom row: 
      efficiencies for different combinations of observables  (red: $(x_E, g)$, blue:
      $(x_E, n_{{\rm ch}}$), green: $(g, n_{{\rm ch}})$). The displayed results refer 
      to charged tracks with minimum $p_\perp^{{\rm tr}}$ of 1~GeV. }
\label{fig:Efficiencies1}}
\end{figure}

\begin{figure}[th!]
\includegraphics[scale=0.49]{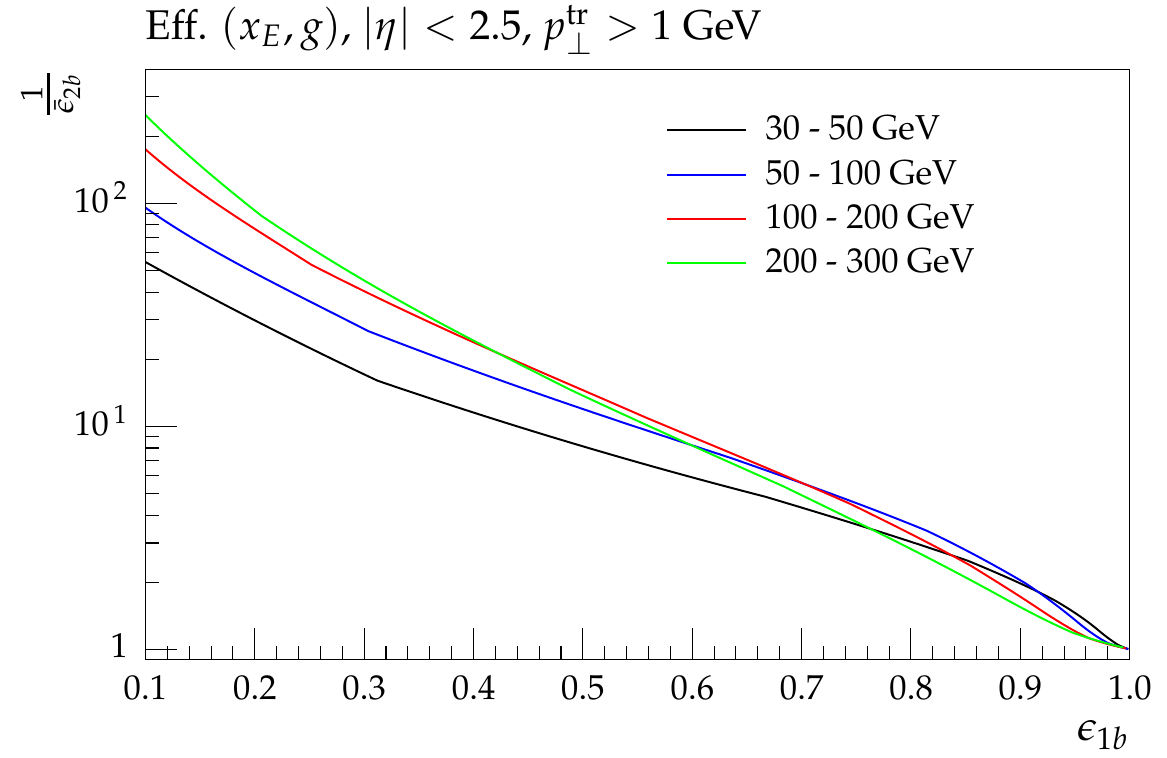}\hspace{-0.2cm}
\includegraphics[scale=0.49]{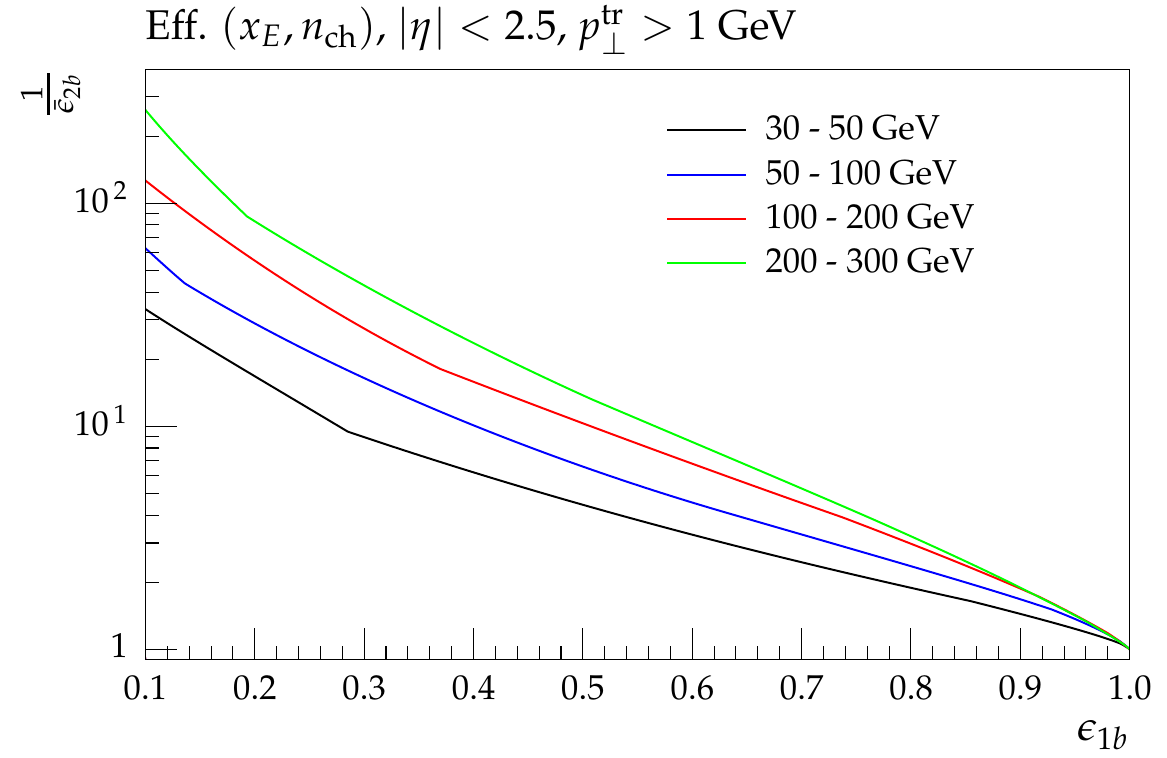}\hspace{-0.2cm}
\includegraphics[scale=0.49]{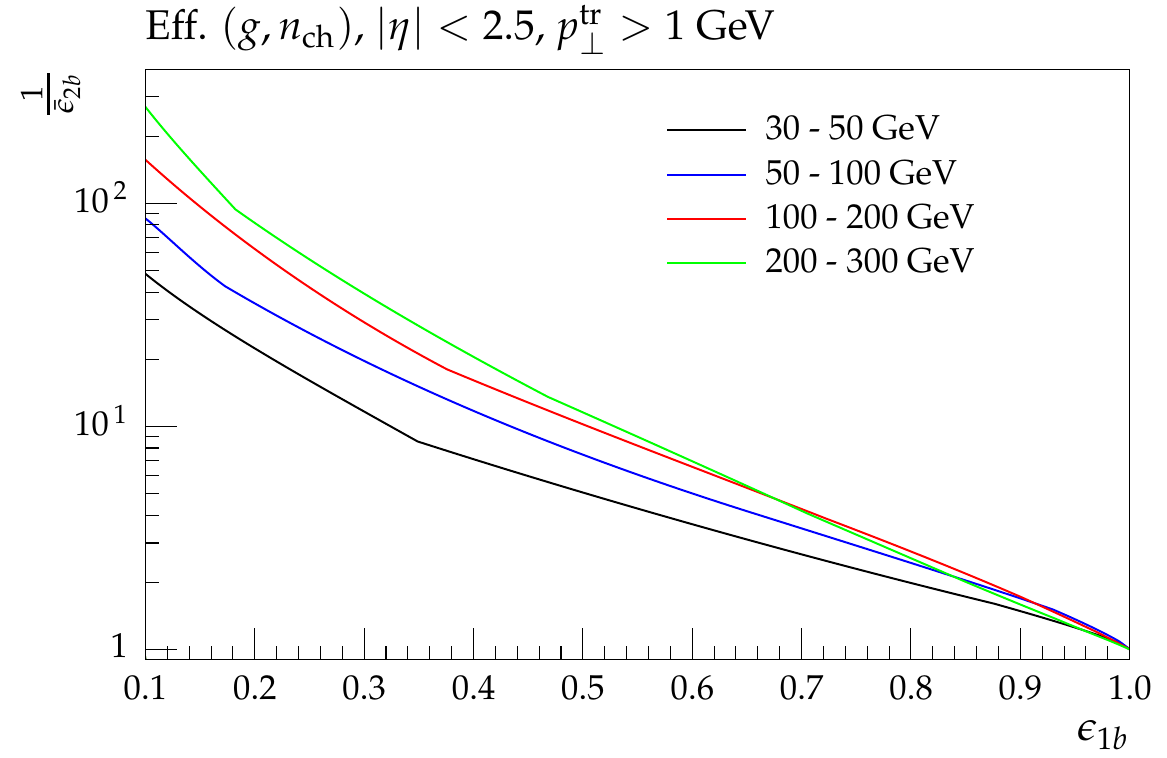}\hspace{-0.2cm}
\parbox{0.8\textwidth}{
  \caption{Efficiency for tagging a b-jet as containing one $b$-hadron
    $\epsilon_{1b}$ against the rejection of jets containing two $b$-hadrons
    $1/\bar{\epsilon}_{2b}$ as a function of $p_{\perp}^J$ for different combinations
    of observables. Left: $(x_E, g)$, centre: $(x_E, n_{{\rm ch}}$), right: 
    $(n_{{\rm ch}},g)$.}
\label{fig:Efficiencies2}}
\end{figure}

The observables  $x_E$, $g$ and $n_{{\rm ch}}$ provide good sensitivity towards the
single and double $b$-tagging samples when considered independently.  As
most of the $b$-tagging algorithms resort to Multivariate Analysis (MVA) with
the combination of the most significative distributions, it is important to
assure that these observables do not present the same correlation pattern and could
therefore generate improved constraints through their combination.  In
Fig.~\ref{fig:Correlation}, the 2-dimensional correlations between the
fragmentation fraction, girth and charged track multiplicity are displayed,
showing only the case for charged tracks of $p_\perp^{{\rm tr}}>1$~GeV for jets
in the $p_\perp^J$-bin between 50 and 100~GeV.  The behaviour seen in these plots
is qualitatively observed also for higher transverse momenta.

Tagging efficiencies are defined based on the so-called ROC curve that uses a simple cut argument. 
For the 1-dimensional distributions, as shown in Fig.~\ref{fig:EFragmentation1}, the  efficiency curve is obtained by 
sliding a cut along the value of the observable. Each point of the cut leads to a correspondent efficiency 
for keeping $b$-jets $(\epsilon_{1b})$ and  $b\bar{b}$-jets $(\epsilon_{2b})$. The ROC curve is the
interpolation of all possible cuts.  For instance, for the girth and multiplicity, a jet is tagged as containing two b-hadrons,
if the value of the observable is above the cut. Whereas, for the fragmentation fractions it is tagged as such 
when the observable is below the cut. The efficiency to tag a jet containing one b-hadron is defined analogously.
The region $x_E < 0.3$ is removed to avoid the lower peak in the boosted regime to maximize the performance of this method.
This region could be efficiently included via a MVA, but this was not done in this letter to retain the simplicity of our strategy.  
The generalization to the combination of two  observables is straightforward: Carrying out a Principal Component Analysis 
on the correlation plots for two $b$-hadrons in the jet, a cut line can be defined perpendicular to the largest eigenvector of 
the correlation. The jet is tagged as containing two $b$-hadrons if  the pair $(x_E, g)$, $(x_E, n_{{\rm ch}})$ or 
$(g, n_{{\rm ch}})$ is above this line. Sliding the cut line along the eigenvector of the correlation matrix, tag and mistag 
efficiencies can be determined.  

These efficiencies are shown in Fig.~\ref{fig:Efficiencies1} (top row) as the efficiency of tagging a $b$-jet
as a jet containing two $b$-hadrons, $\epsilon_{2b}$,  against the rejection of $b$-jets,  $1/\bar{\epsilon}_{1b}$.  
The combination of observables proves to be robust against the choice of charged tracks or the fully hadronic
final state.  Lowering the threshold $p_\perp^{{\rm tr}}$ to 0.5~GeV produces only mild improvements in respect to 1~GeV. 

In Fig.~\ref{fig:Efficiencies1} (bottom row), different combinations of observables are
compared with the discrimination from $n_{{\rm ch}}$ or $x_E$  only.  A sizable improvement in
using the combination of two observables is found.   For low transverse momenta
the combination $(x_E, g)$ outperforms the other combinations, while
for larger transverse momenta of the jet, the combination of $(x_E, n_{{\rm ch}})$
is most sensitive.  In both cases, however, the fragmentation fraction is
involved, an observable that hitherto has not been documented for this
discrimination. In Fig.~\ref{fig:Efficiencies2}, the different combinations are displayed  for 
distinct transverse momenta slices. The $b\bar{b}$-jet rejection efficiency ($1/\bar{\epsilon}_{2b}$) 
significantly improves for the phenomenologically interesting boosted topologies in all cases.
The $(x_E,g)$ produces robust results through all the transverse momentum slices. This 
suggests that the combination of these two observables contains complementary and relevant 
information not found in the single observables or the other combinations.

\section{Summary}
\label{sec:summary}

Studies that require multiple $b$-jets will become increasingly
frequent at the LHC in the years to come.  These studies range from SM
precision analyses to searches for beyond the SM physics, such as resonance
searches.  One of the problems encountered is related to discriminating
the ``legitimate'' $b$-jets, containing only one, typically hard $b$-hadron,
from jets containing two $b$-hadrons, usually emerging from a gluon splitting.
In this publication a phenomenological attempt at a more coherent strategy of 
discriminating $b$ and $b\bar{b}$-jets has been presented, based on possible
kinematic handles, in particular combinations of jet shapes with the fragmentation 
fraction.

Several observables were considered and the most powerful encountered were
the girth $g$, the number of charged tracks $n_{{\rm ch}}$, and $b$-hadron jet
energy fraction $x_E$.  Especially when combining either of the former two with
the latter a considerable improvement was found. A significant improvement for the 
$b\bar{b}$-jet rejection is observed at the boosted regime for all variable combinations.

\section*{Acknowledgements}
The authors acknowledge financial support by the  UK  Science and Technology
Facilities Council, by the Research Executive Agency (REA) of the European
Union under the Grant Agreements PITN-GA-2012-316704 (``HiggsTools")
and PITN-GA-2012-315877 (``MCnetITN''), and by the ERC Advanced Grant
MC@NNLO (340983).

\bibliography{b-Jets}{}
\bibliographystyle{unsrt}
\end{document}